\documentclass[10pt,conference,letterpaper]{IEEEtran}
\pdfoutput=1
\usepackage{amsmath,amssymb,graphicx}
\usepackage[linesnumbered,ruled,vlined]{algorithm2e}
\usepackage{balance}
\usepackage[caption=false,font=footnotesize]{subfig}
\usepackage[hyphens]{url}
\usepackage[bookmarks=true,hidelinks]{hyperref}
\usepackage{hypcap}
\usepackage{xcolor}
\hypersetup{
	colorlinks,
	linkcolor={red!50!black},
	citecolor={blue!50!black},
	urlcolor={blue!80!black}
}

\newtheorem{theorem}{Theorem}
\newtheorem{definition}{Definition}
\newtheorem{example}{Example}
\newtheorem{lemma}{Lemma}

\newtheorem{corollary}{Corollary}

\graphicspath{{figures/}}

\author{
	\IEEEauthorblockN{Jing Tang}
	\IEEEauthorblockA{School of Comp. Sci. \& Eng.\\
		Nanyang Technological University\\
		Email: tang0311@ntu.edu.sg}
	\and
	\IEEEauthorblockN{Xueyan Tang}
	\IEEEauthorblockA{School of Comp. Sci. \& Eng.\\
		Nanyang Technological University\\
		Email: asxytang@ntu.edu.sg}
	\and
	\IEEEauthorblockN{Junsong Yuan}
	\IEEEauthorblockA{School of EEE\\
		Nanyang Technological University\\
		Email: jsyuan@ntu.edu.sg}
}

\title{Towards Profit Maximization for Online Social Network Providers}
\begin{document}
\maketitle
\begin{abstract}
Online Social Networks (OSNs) attract billions of users to share information and communicate where viral marketing has emerged as a new way to promote the sales of products. An OSN provider is often hired by an advertiser to conduct viral marketing campaigns. The OSN provider generates revenue from the commission paid by the advertiser which is determined by the spread of its product information. Meanwhile, to propagate influence, the activities performed by users such as viewing video ads normally induce diffusion cost to the OSN provider. In this paper, we aim to find a seed set to optimize a new profit metric that combines the benefit of influence spread with the cost of influence propagation for the OSN provider. Under many diffusion models, our profit metric is the difference between two submodular functions which is challenging to optimize as it is neither submodular nor monotone. We design a general two-phase framework to select seeds for profit maximization and develop several bounds to measure the quality of the seed set constructed. Experimental results with real OSN datasets show that our approach can achieve high approximation guarantees and significantly outperform the baseline algorithms, including state-of-the-art influence maximization algorithms.
\end{abstract}

\section{Introduction}\label{sec:introduction}
Information can be disseminated widely and rapidly through Online Social Networks (OSNs) (such as Facebook, Twitter, Flickr, Google+, and LinkedIn) with ``word-of-mouth" effects. Viral marketing is a typical application that leverages OSNs as the medium for information diffusion \cite{Domingos_maxInfluence_2001}. The market of OSN advertisement is growing explosively. For example, Fortune \cite{Roberts_socialAds_2016} reports that the online and digital advertisement spending for the 2016 election of the United States will reach $1.2$ billion US dollars in which $49\%$ is expected to go to social media.

Advertisers often delegate the operation of viral marketing campaigns to OSN providers who have the complete information of their social network structures \cite{Chalermsook_SNM_2015}. Providing viral marketing services is a potential and promising approach that OSN providers can explore for monetization. When hiring an OSN provider to conduct the viral marketing campaign, the advertiser usually pays the OSN provider a commission for each user that adopts its product or shares its ads. Thus, the revenue of the OSN provider generated from the commission is determined by the spread of the product information. Meanwhile, to propagate influence, the activities performed by users such as viewing video ads normally induce diffusion cost to the OSN provider. For example, for a cloud-based OSN running viral video marketing \cite{Meyer_VVM_2015}, the OSN will be charged by the cloud for each video click due to the data traffic produced. Intuitively, the number of clicks for each video is dependent on the number of connections among users that are actually used during the viral marketing campaign. Therefore, to maximize its profit, the OSN provider needs to account for both the reward of influence spread and the expense of influence propagation, both of which are dependent on the seed users selected to initialize the viral marketing campaign. In this paper, we study the problem of finding a seed set to optimize such a new profit metric that combines the benefit of influence spread with the cost of influence propagation.

Our profit metric is challenging to optimize as it is significantly different from the influence metric that has been studied widely \cite{Chen_MIA_2010,Chen_degreeDiscount_2009,Goyal_infMax_2011,Jung_IRIE_2012,Kempe_maxInfluence_2003,Leskovec_CELF_2007,Nguyen_BCT_2016,Nguyen_DSSA_2016,Ohsaka_prunedMC_2014,Song_CGA_2015,Tang_OPIM_2018,Tang_infMax_2017,Tang_IMM_2015,Tang_reverse_2014,Zhou_UBLF_2013}. It is well known that the influence spread generated by a seed set is submodular and monotone under many diffusion models \cite{Barbieri_TAI_2012,Kempe_maxInfluence_2003,Rodriguez_CT_2011}, which makes it easy to design influence maximization algorithms with strong approximation guarantees \cite{Nemhauser_submodular_1978}. Some recent studies have addressed profit maximization from the advertiser's perspective by putting together the benefit of influence spread and the cost of seed selection \cite{Lu_maxprofit_2012,Tang_profitMax_2016,Tang_profitMaxUS_2018,Zhu_maxprofit_2013}. In these studies, the cost of seed selection is modeled by the incentives (e.g., free samples) provided to seed users. Since the seed selection cost is given by the sum of the costs of individual seeds which is modular, the resultant profit metric is still submodular. In contrast, in our problem, both the revenue and cost of the OSN provider are attached to the diffusion process. As a result, our profit metric can be viewed as the difference between two submodular functions, which is neither submodular nor monotone and is thus much more difficult to deal with. 

In this paper, we propose a general two-phase framework to optimize profit for the OSN provider. Our framework is comprised of a pruning phase that iteratively narrows down the search space and a search phase that uses heuristics to select seed nodes within the reduced search space. We theoretically establish the advantages of our pruning technique in improving the effectiveness and efficiency of any algorithm used in the search phase. We further derive several bounds on the maximum achievable profit to evaluate the quality of the seed sets obtained by any algorithm. We conduct extensive experiments with several real OSN datasets. The results demonstrate the effectiveness and efficiency of our framework.

The rest of this paper is organized as follows. Section \ref{sec:relatedWork} reviews the related work. Section \ref{sec:problem} defines the profit maximization problem for the OSN provider. Section \ref{sec:algorithms} elaborates the design of our two-phase framework. Section \ref{sec:analysis} develops the bounds for quality measurement. Section \ref{sec:evaluation} presents the experimental evaluation. Finally, Section \ref{sec:conclusion} concludes the paper.

\section{Related Work}\label{sec:relatedWork}

\textbf{Influence Maximization.} 
Kempe et al. \cite{Kempe_maxInfluence_2003} formulated an influence maximization problem for a given seed set size with two basic diffusion models, namely the Independent Cascade (IC) and Linear Threshold (LT) models. They showed that the influence spreads under these models are submodular and monotone. Thus, they proposed a simple hill-climbing greedy algorithm to address the problem, which can provide a $(1-1/e)$-approximation guarantee \cite{Nemhauser_submodular_1978}. The follow-up studies have mainly concentrated on improving the efficiency of the algorithm implementation for large-scale OSNs \cite{Chen_MIA_2010,Chen_degreeDiscount_2009,Goyal_infMax_2011,Jung_IRIE_2012,Leskovec_CELF_2007,Nguyen_BCT_2016,Nguyen_DSSA_2016,Ohsaka_prunedMC_2014,Song_CGA_2015,Tang_OPIM_2018,Tang_infMax_2017,Tang_IMM_2015,Tang_reverse_2014,Zhou_UBLF_2013}. Different from the above studies, we target at maximizing the profit that accounts for both the benefit of influence spread and the cost of influence propagation in viral marketing.

\textbf{Profit Maximization.} 
Maximizing the influence spread alone has been shown to be ineffective for optimizing the profit return of viral marketing \cite{Tang_profitMax_2016,Tang_profitMaxUS_2018}. This is because the number of seeds selected yields a tradeoff between the expense and reward of viral marketing. To avoid pre-setting the number of seeds to select, some recent work studied profit maximization from the advertiser's perspective \cite{Lu_maxprofit_2012,Tang_profitMax_2016,Tang_profitMaxUS_2018,Zhu_maxprofit_2013}. These studies considered the cost of seed selection which is modular and implies that their profit metric is still submodular. They investigated heuristics to select seeds under the assumption that the social network structures are available to the advertiser. In practice, only the OSN providers have the complete information of their social graphs and they often keep the formation secret for business and privacy reasons \cite{Chalermsook_SNM_2015,Khan_revmax_2016}. Therefore, the OSN providers are able to run viral marketing campaigns more efficiently than the advertisers. Different from the above work, we formulate a new profit maximization problem from the OSN provider's perspective that takes into account the cost of information diffusion over the social network. As shall be shown later, our profit metric is the difference between two submodular functions, which is neither submodular nor monotone. 

\textbf{Submodular Optimization.}
Iyer and Bilmes \cite{Iyer_submodularDiff_2012} investigated optimizing the difference between two submodular functions. They showed that this problem is multiplicatively inapproximable unless P=NP and proposed several heuristic methods to address the problem. The greedy algorithm \cite{Kempe_maxInfluence_2003,Tang_profitMax_2016} is another widely used heuristic for submodular optimization. We apply these heuristic methods to the search phase of our two-phase framework. More importantly, we develop an iterative pruning technique to reduce the search space which significantly improves these heuristics in terms of not only effectiveness but also efficiency. In addition, we also derive several upper bounds of the optimum to benchmark the output of any algorithm.

\section{Problem Formulation}\label{sec:problem}
\subsection{Preliminaries}\label{subsec:ICmodel}
Let ${G}=({V},{E})$ be a directed graph modeling an OSN, where the nodes ${V}$ represent users and the edges ${E}$ represent the connections among users (e.g., friendships on Facebook, followships on Twitter). For each directed edge $\langle{u,v}\rangle\in {E}$, we refer to $v$ as a \emph{neighbor} of $u$, and refer to $u$ as an \emph{inverse neighbor} of $v$.

There are many diffusion models for the processes by which influence propagates in social networks. The diffusion model is normally a random process that starts with a set of seed nodes ${S}$. Initially, the seed nodes ${S}$ are activated, while all the other nodes are not activated. When a node $u$ becomes activated, it would attempt to further activate its neighbors which are not yet activated. The diffusion process terminates when no more node can be further activated. Let ${g}\sim{G}$ be a sample outcome of influence propagation by the diffusion process and let ${V}_{g}({S})$ be the set of nodes activated by the seed set ${S}$ in the sample outcome ${g}$. The \textit{influence spread} of the seed set ${S}$, denoted by $\sigma({S})$, is the expected number of nodes activated over all possible sample outcomes of influence propagation, i.e., $\sigma({S})=\mathbb{E}[|{V}_{g}({S})|]$.

\subsection{The Profit Maximization Problem}\label{subsec:problem}
As discussed, the influence spread is the benefit gained by the OSN provider and the cost of influence propagation is the price to pay for viral marketing. We assume that each node $v$ in the social network is associated with a benefit weight $b(v)$, which represents the benefit offered by activating $v$ (e.g., the commission received from the advertiser).
The benefit $\beta({S})$ of influence spread generated by a seed set ${S}$ is the total benefit brought by all the nodes activated:
\begin{equation}\label{eq:benefit}
\beta({S})=\mathbb{E}\Big[\sum_{v\in{V}_{g}({S})}b(v)\Big].
\end{equation}
To model the cost of influence propagation, we consider the diffusion process in the social network. Recall that when a node is activated, it attempts to further activate its neighbors through the connections to them in the social network. Without loss of generality, we assume that each node $v$ is associated with a cost weight $c(v)$, which represents the cost of information diffusion incurred by $v$ in attempting to activate its neighbors when it becomes activated (e.g., pushing video ads). The cost $\gamma({S})$ of influence propagation introduced by a seed set ${S}$ is the total cost incurred by all the nodes activated: 
\begin{equation}\label{eq:cost}
\gamma({S})=\mathbb{E}\Big[\sum_{v\in{V}_{g}({S})}c(v)\Big].
\end{equation}
Then, we naturally define a \textit{profit} metric from OSN provider's perspective as the benefit of influence spread less the cost of influence propagation, i.e., the profit $\phi({S})$ for running a viral marketing campaign with a seed set ${S}$ is given by
\begin{equation}\label{eq:profit}
\phi({S})=\beta({S})-\gamma({S}),
\end{equation}
Our goal is to find a seed set ${S}$ to maximize the profit $\phi({S})$.

By defining $w(v)=b(v)-c(v)$, we can rewrite the profit metric in \eqref{eq:profit} as
\begin{equation*}
\phi({S})=\mathbb{E}\Big[\!\!\sum_{v\in{V}_{g}({S})}\!\!\big(b(v)-c(v)\big)\Big]\triangleq\mathbb{E}\Big[\!\!\sum_{v\in{V}_{g}({S})}\!\!w(v)\Big].
\end{equation*}
Here, $w(v)$ represents the profit gain of activating a node $v$. If the benefit offered by $v$ outweighs its cost, $w(v)$ is positive. Otherwise, $w(v)$ is negative. Therefore, we can normalize the benefit and cost weights of each node $v$ by setting $\bar{b}(v)=\max\{0,w(v)\}$ and $\bar{c}(v)=\max\{0,-w(v)\}$. Then, the benefit and cost metrics become 
\begin{equation*}
\bar{\beta}({S})=\mathbb{E}\Big[\sum_{v\in{V}_{g}({S})}\bar{b}(v)\Big],
\text{\quad and\quad}\bar{\gamma}({S})=\mathbb{E}\Big[\sum_{v\in{V}_{g}({S})}\bar{c}(v)\Big],
\end{equation*}
respectively. We show later that this normalized form can make the pruning phase of our proposed framework more effective.

\begin{figure}[!t]
	\capstart
	\centering
	\includegraphics[width=1.0\linewidth]{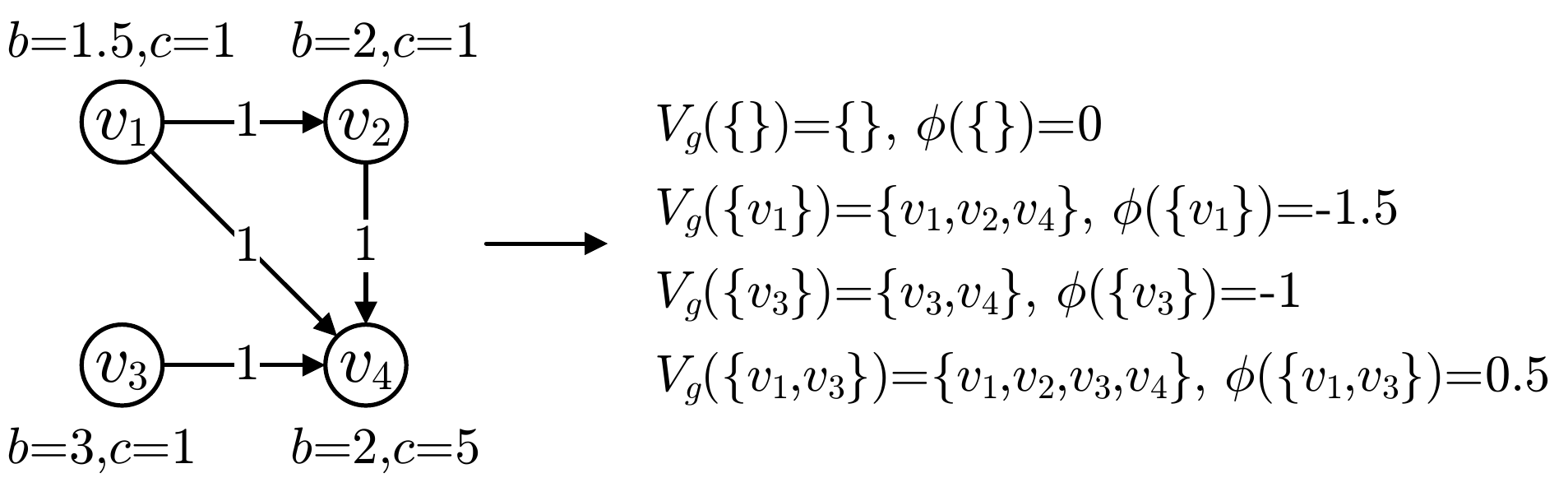}
	\vspace{-0.15in}
	\caption{An example showing the non-monotonicity and non-submodularity of the profit function $\phi(\cdot)$. Each node is associated with a benefit weight $b$ and a cost weight $c$.}\label{fig:example_objective}
	\vspace{-0.1in}
\end{figure}

It has been proved that the influence function $\sigma(\cdot)$ is submodular under many commonly used diffusion models due to ${V}_{g}({S})\subseteq {V}_{g}({T})$ for any ${S}\subseteq {T}$ \cite{Kempe_maxInfluence_2003}. Similarly, it can be shown that the benefit and cost metrics defined above are both submodular under these diffusion models. That is, for any two seed sets ${S}$ and ${T}$ where ${S}\subseteq{T}$ and any node $v\notin{T}$, it holds that $\beta({S}\cup\{v\})-\beta({S})\geq\beta({T}\cup\{v\})-\beta({T})$ and $\gamma({S}\cup\{v\})-\gamma({S})\geq\gamma({T}\cup\{v\})-\gamma({T})$. Likewise, the normalized $\bar{\beta}(\cdot)$ and $\bar{\gamma}(\cdot)$ are also submodular. Although both the benefit and cost metrics are monotone and submodular, the profit metric is neither monotone nor submodular (an example is given in \figurename~\ref{fig:example_objective}). Thus, the methods used for maximizing monotone submodular functions would perform poorly for our profit maximization problem as shall be shown in our experiments. In the following, we develop heuristic algorithms to address our profit maximization problem.

\begin{example}
	\figurename~\ref{fig:example_objective} gives an example to illustrate that the profit function $\phi(\cdot)$ is neither monotone nor submodular. In this example, once a node is activated, it will activate all of its neighbors. It can be seen from the figure that $\phi(\emptyset)=0$, $\phi(\{v_1\})=-1.5$, $\phi(\{v_3\})=-1$ and, $\phi(\{v_1,v_3\})=0.5$. Thus, $\phi(\emptyset)>\phi(\{v_1\})$ and $\phi(\{v_1\})<\phi(\{v_1,v_3\})$, which indicates that $\phi(\cdot)$ is non-monotone. In addition, we can also get that $\phi(\{v_1\})-\phi(\emptyset)=-1.5 < \phi(\{v_1,v_3\})-\phi(\{v_3\})=1.5$, which indicates that $\phi(\cdot)$ is non-submodular.
\end{example}

\section{Two-Phase Framework}\label{sec:algorithms}
We propose a general two-phase framework to select seed nodes for optimizing the profit.
\begin{itemize}
	\item \textbf{Pruning phase (Algorithm~\ref{alg:IterativePrune}):} We develop an iterative pruning technique to cut the search space.
	\item \textbf{Search phase (Algorithms~\ref{alg:greedy} and \ref{alg:MM}):} We use some heuristics to find the solution in the reduced search space.
\end{itemize}

\subsection{Prune Search Space}\label{subsection:pruning}
To simplify the notations, we define the marginal gain of adding a node $v$ to a seed set ${S}$ for metrics $\phi$, $\beta$ and $\gamma$ as
\begin{align*}
&\phi(v\mid{S})=\phi({S}\cup\{v\})-\phi({S}),\\
&\beta(v\mid{S})=\beta({S}\cup\{v\})-\beta({S}),\\
&\gamma(v\mid{S})=\gamma({S}\cup\{v\})-\gamma({S}).
\end{align*}

\begin{algorithm}[!t]
	\begin{small}
		\caption{IterativePrune}
		\label{alg:IterativePrune}
		start with ${A}_0\leftarrow \emptyset$, ${B}_0\leftarrow {V}$ and $t=0$\;\label{alg:IterativePrune_init}
		\Repeat{converged, i.e., ${A}_t={A}_{t-1}$ and ${B}_t={B}_{t-1}$\label{alg:IterativePrune_stop}}{
			${A}_{t+1}\leftarrow {A}_t\cup\{v\colon\beta(v\mid{B}_t\setminus\{v\})-\gamma(v\mid{A}_t)>0$ and $v\in{B}_t\setminus{A}_t\}$\;\label{alg:IterativePrune_At}
			${B}_{t+1}\leftarrow {B}_t\setminus\{v\colon\beta(v\mid{A}_t)-\gamma(v\mid{B}_t\setminus\{v\})<0$ and $v\in{B}_t\setminus{A}_t\}$\;\label{alg:IterativePrune_Bt}
			$t\leftarrow t+1$\;\label{alg:IterativePrune_t}
		}
		\Return ${A}_t$ and ${B}_t$ as ${A}^\ast$ and ${B}^\ast$\;\label{alg:IterativePrune_return}
	\end{small}
\end{algorithm}

We start by proposing an iterative pruning approach that can dramatically reduce the search space from the power set of ${V}$ to a smaller lattice for maximizing the profit function. Algorithm~\ref{alg:IterativePrune} shows the pseudo code of the pruning algorithm. Recall that by the submodularity of the benefit and cost metrics, their marginal gains for adding a new seed node decrease with the base seed set. Thus, the largest possible benefit (cost) gain is produced by adding a node into an empty seed set, whereas the smallest possible benefit (cost) gain is generated by adding the node into an almost universal set. Note that the marginal profit gain is bounded below by the smallest benefit gain less the largest cost gain. So, it is intuitive that if the latter is positive, the node must be selected in an optimal solution. Similarly, the marginal profit gain is bounded above by the largest benefit gain less the smallest cost gain. If the latter is negative, the node cannot be selected in an optimal solution. Algorithm~\ref{alg:IterativePrune} extends this idea in an iterative manner to reduce the search space. It is easy to verify that after each iteration, the newly generated lattice is a sublattice of that in the previous iteration, i.e., ${A}_{t}\subseteq{A}_{t+1}\subseteq{B}_{t+1}\subseteq{B}_{t}$. Furthermore, it can be proved that any seed set outside the resultant lattice $[{A}^\ast,{B}^\ast]$ delimited by the node sets ${A}^\ast$ and ${B}^\ast$ returned by Algorithm~\ref{alg:IterativePrune} can be transformed to a seed set in $[{A}^\ast,{B}^\ast]$ with higher profit. The formal proofs of all the theoretical results in this paper are given in the appendix.
\begin{theorem}\label{theorem:pruneImprovement}
	For any node set ${S}$ and $t\geq 0$, let ${S}_t={S}\cap{B}_t\cup{A}_t$, it holds that $\phi({S}_{t})\leq\phi({S}_{t+1})$,
	where the ``$=$'' holds if and only if ${S}_{t}={S}_{t+1}$.
\end{theorem}

Theorem~\ref{theorem:pruneImprovement} shows that the pruning technique can only increase the profit value of the seed set ${S}_t={S}\cap{B}_t\cup{A}_t$ at every iteration. The following corollary shows how our search space reduction improves the quality of any solution.
\begin{corollary}\label{corollary:pruneImprovement}
	For any node set ${S}$, if ${S}\notin[{A}^\ast,{B}^\ast]$, then $\phi({S})<\phi({S}\cap{B}^\ast\cup{A}^\ast)$.
\end{corollary}

By Corollary~\ref{corollary:pruneImprovement}, the pruning approach can always improve the quality of any seed set ${S}$ outside $[{A}^\ast,{B}^\ast]$ by transforming it to the seed set ${S}\cap{B}^\ast\cup{A}^\ast$ in $[{A}^\ast,{B}^\ast]$. Thus, the lattice $[{A}^\ast,{B}^\ast]$ retains all the optimal seed sets. As a result, we can reduce the search space from the lattice $[\emptyset,{V}]$ to $[{A}^\ast,{B}^\ast]$. Our pruning approach can be used prior to any seed selection algorithms to improve the solution quality.
\begin{corollary}
	\label{corollary:allMaximizers}
	For any seed set ${S}^\ast$ producing the maximum achievable profit, it holds that ${A}^\ast\subseteq{S}^\ast\subseteq{B}^\ast$. 
\end{corollary}

\begin{figure}[!t]
	\capstart
	\centering
	\includegraphics[width=1.0\linewidth]{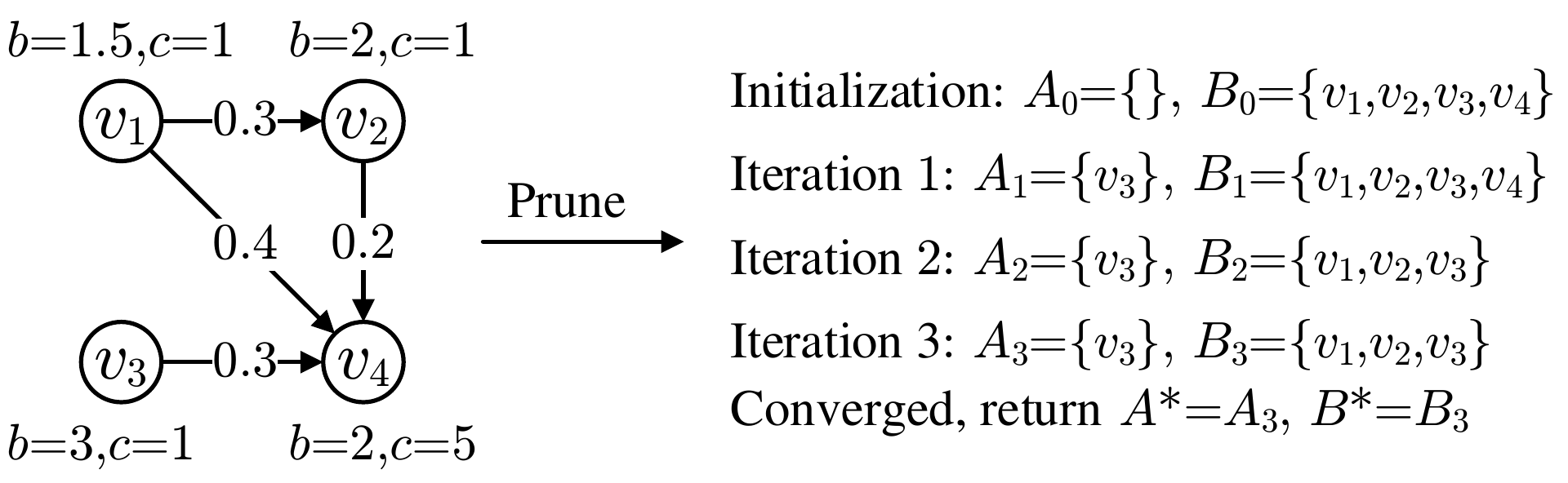}
	\vspace{-0.2in}
	\caption{An example of iterative pruning under the Independent Cascade diffusion model. Each node is associated with a benefit weight $b$ and a cost weight $c$. Each edge has a propagation probability $p$.}\label{fig:example_prune}
	\vspace{-0.15in}
\end{figure}

\begin{example}\label{example:pruning}
	\figurename~\ref{fig:example_prune} gives an example to illustrate how the pruning algorithm works as well as the above theorem and corollaries. This example assumes the Independent Cascade (IC) diffusion model. The IC model is a representative and most widely-studied diffusion model for influence propagation \cite{Chen_MIA_2010,Chen_degreeDiscount_2009,Jung_IRIE_2012,Kempe_maxInfluence_2003,Leskovec_CELF_2007,Nguyen_BCT_2016,Nguyen_DSSA_2016,Ohsaka_prunedMC_2014,Song_CGA_2015,Tang_OPIM_2018,Tang_infMax_2017,Tang_IMM_2015,Tang_reverse_2014,Zhou_UBLF_2013}. In the IC model, a propagation probability $p_{u,v}$ is associated with each edge $\langle{u,v}\rangle$, representing the probability for $v$ to be activated by $u$ through their connection. In the diffusion process, when a node $u$ first becomes activated, it has a chance to activate its neighbors who are not yet activated. Each such neighbor $v$ would become activated with probability $p_{u,v}$. This process repeats until no more node can be activated. For example, in \figurename~\ref{fig:example_prune}, when $\{v_1,v_3\}$ are selected as seeds, $v_2$ would be activated with probability $0.3$. Meanwhile, $v_4$ would be activated by $v_1$ with probability $0.4$, by $v_2$ with probability $0.3\times 0.2=0.06$, and by $v_3$ with probability $0.3$. Thus, overall, $v_4$ would be activated  with probability $1-(1-0.4)\times(1-0.06)\times(1-0.3)=0.6052$.
	
	To conduct iterative pruning, ${A}_0$ is initialized by $\emptyset$ and ${B}_0$ is initialized by $\{v_1,v_2,v_3,v_4\}$ respectively. To simplify the notations, let $\phi^-_t(v)=\beta(v\mid{B}_t\setminus\{v\})-\gamma(v\mid{A}_t)$ and $\phi^+_t(v)=\beta(v\mid{A}_t)-\gamma(v\mid{B}_t\setminus\{v\})$ describe the calculations in Algorithm~\ref{alg:IterativePrune}. At iteration 1, $\phi^-_1(v_1)=\beta(v_1\mid\{v_2,v_3,v_4\})-\gamma(v_1\mid\emptyset)=1\times b(v_1)-\big(1\times c(v_1)+0.3\times c(v_2)+(1-(1-0.4)\times(1-0.3\times 0.2))\times c(v_4)\big)=1.5-(1+0.3+2.18)=1.5-3.48=-1.98<0$. Similarly, $\phi^-_1(v_2)=-0.6<0$, $\boldsymbol{\phi^-_1(v_3)=0.5>0}$, $\phi^-_1(v_4)=-4.328<0$, $\phi^+_1(v_1)=1.972>0$, $\phi^+_1(v_2)=1.7>0$, $\phi^+_1(v_3)=2.6>0$, and $\phi^+_1(v_4)=0.32>0$. Thus, $v_3$ is added to ${A}_1$ so that ${A}_1=\{v_3\}$ and ${B}_1=\{v_1,v_2,v_3,v_4\}$. At iteration 2, $\phi^-_2(v_1)=-1.326<0$, $\phi^-_2(v_2)=-0.3<0$, $\phi^-_2(v_4)=-2.828<0$, $\phi^+_2(v_1)=1.7104>0$, $\phi^+_2(v_2)=1.58>0$, and $\boldsymbol{\phi^+_2(v_4)=-0.28<0}$. Thus, $v_4$ is removed from ${B}_2$ so that ${A}_2=\{v_3\}$ and ${B}_2=\{v_1,v_2,v_3\}$. At iteration 3, $\phi^-_3(v_1)=-0.878<0$, $\phi^-_3(v_2)=-0.1824<0$, $\phi^+_3(v_1)=0.5904>0$, and $\phi^+_3(v_2)=1.286>0$. Thus, both ${A}_3$ and ${B}_3$ remain the same as in the previous iteration. As a result, ${A}^\ast=\{v_3\}$ and ${B}^\ast=\{v_1,v_2,v_3\}$ are returned. For a seed set ${S}=\{v_2,v_4\}\notin[{A}^\ast,{B}^\ast]$, we have ${S}_1={S}\cap{B}_1\cup{A}_1=\{v_2,v_3,v_4\}$ and ${S}_2={S}_3={S}\cap{B}^\ast\cup{A}^\ast=\{v_2,v_3\}$. Then, it can be obtained that $\phi({S})=-2<\phi({S}_1)=0<\phi({S}_2)=1.68$, which demonstrates Theorem~\ref{theorem:pruneImprovement} and Corollary~\ref{corollary:pruneImprovement}. Moreover, it is easy to verify that the optimal seed set ${S}^\ast=\{v_2,v_3\}$ belongs to $[{A}^\ast,{B}^\ast]$, which confirms Corollary~\ref{corollary:allMaximizers}.
\end{example}

Finally, it can be shown that normalizing the benefit and cost weights as described in Section~\ref{subsec:problem} can only increase the amount of the search space cut by our pruning technique.

\begin{theorem}\label{theorem:normalization_prune}
	Let $\bar{{A}}^\ast$ and $\bar{{B}}^\ast$ be the node sets returned by Algorithm~\ref{alg:IterativePrune} under the normalized form. Then, ${A}^\ast\subseteq\bar{{A}}^\ast\subseteq\bar{{B}}^\ast\subseteq{B}^\ast$.
\end{theorem}

We shall experimentally evaluate the additional reduction in the search space due to normalization in Section~\ref{sec:evaluation}.

\subsection{Heuristic Algorithms}
We now present some heuristic methods to address the profit maximization problem since there does not exist any polynomial time algorithm with any polynomial time multiplicative approximation guarantees unless P=NP \cite{Iyer_submodularDiff_2012}.

\textbf{Greedy Algorithm:} 
We apply a simple hill-climbing idea to optimize the profit function \eqref{eq:profit}. Algorithm~\ref{alg:greedy} describes the pseudo code. In each iteration, the greedy heuristic adds a new node $u$ to ${S}$ that has the largest marginal profit gain $\phi(u\mid{S})$ until all the remaining nodes have negative marginal gains.

\begin{algorithm}[!h]
	\begin{small}
		\caption{Greedy}
		\label{alg:greedy}
		initialize ${S}\leftarrow {A}^\ast$\;\label{alg:greedy_init}
		\While{True}{
			find $u\leftarrow\arg\max_{v\in {B}^\ast\setminus {S}}\left\{\phi(v\mid{S})\right\}$\;\label{alg:greedy_findu}
			\lIf{$\phi(u\mid{S})\leq0$}{\Return ${S}$\label{alg:greedy_return}}
			${S}\leftarrow {S}\cup \{u\}$\;\label{alg:greedy_addone}
		}
	\end{small}
\end{algorithm}

\textbf{Modular-Modular Algorithm:} 
Iyer and Bilmes \cite{Iyer_submodularDiff_2012} introduced a modular-modular (ModMod) algorithm for optimizing the difference between submodular functions. Since we just need to search the lattice $[{A}^\ast,{B}^\ast]$ after pruning, we adapt the ModMod algorithm as shown in Algorithm~\ref{alg:MM}. In line~\ref{alg:MM_Xt} of Algorithm~\ref{alg:MM}, $h_{{X}^t}^{\pi}({Y};\beta)$ is a modular lower bound of $\beta({Y})$ that is tight at set ${X}^t$, i.e., $h_{{X}^t}^{\pi}({Y};\beta)\leq \beta({Y})$ for any ${Y}\subseteq{V}$ and $h_{{X}^t}^{\pi}({X}^t;\beta)=\beta({X}^t)$, while $m_{{X}^t}({X}^t;\gamma)$ is a modular upper bound of $\gamma({Y})$ that is also tight at ${X}^t$, i.e., $m_{{X}^t}({Y};\gamma)\geq \gamma({Y})$ for any ${Y}\subseteq{V}$ and $m_{{X}^t}({X}^t;\gamma)=\gamma({X}^t)$. Thus, the difference $h_{{X}^t}^{\pi}({Y};\beta)-m_{{X}^t}({Y};\gamma)$ is a lower bound of the profit function $\phi({Y})$. The algorithm maximizes the lower bound in each iteration. Since the lower bound is tight at ${Y}={X}^t$, it is guaranteed that $\phi({X}^{t+1})\geq h_{{X}^t}^{\pi}({X}^{t+1};\beta)-m_{{X}^t}({X}^{t+1};\gamma) \geq h_{{X}^t}^{\pi}({X}^{t};\beta)-m_{{X}^t}({X}^{t};\gamma)=\phi({X}^{t})$. This indicates that Algorithm~\ref{alg:MM} always increases the profit value at every iteration. Examples of the modular upper and lower bounds will be given in Section~\ref{subsec:online_bounds}.

\begin{algorithm}[!t]
	\begin{small}
		\caption{Modular-Modular (ModMod)}
		\label{alg:MM}
		initialize ${X}^0\leftarrow{A}^\ast$ and $t\leftarrow 0$\;\label{alg:MM_init}
		\Repeat{converged, i.e., ${X}^t={X}^{t-1}$\label{alg:MM_stop}}{
			choose the permutations of ${A}^\ast$, ${X}^t\setminus{A}^\ast$, ${B}^\ast\setminus{X}^t$ and concatenate them as $\pi$\;\label{alg:MM_permutation}
			${X}^{t+1}\leftarrow\arg\max_{{A}^\ast\subseteq{Y}\subseteq{B}^\ast}h_{{X}^t}^{\pi}({Y};\beta)-m_{{X}^t}({Y};\gamma)$\;\label{alg:MM_Xt}
			$t\leftarrow t+1$\;\label{alg:MM_t}
		}
		\Return ${X}^t$\;\label{alg:MM_return}
	\end{small}
\end{algorithm}

\subsection{Discussions}\label{subsec:discussions}
\textbf{Time Complexity:}
Evaluating the profit metric involves estimating the influence spread given a seed set. Any existing influence estimation methods, such as Monte-Carlo simulation \cite{Kempe_maxInfluence_2003,Leskovec_CELF_2007,Ohsaka_prunedMC_2014} and reverse influence sampling \cite{Borgs_RIS_2014,Nguyen_BCT_2016,Nguyen_DSSA_2016,Tang_IMM_2015,Tang_reverse_2014}, can be used. Suppose the time complexity for computing the marginal profit gain of adding/removing a node into/from a seed set is $O(M)$. For the iterative pruning process (Algorithm~\ref{alg:IterativePrune}), the size of the node set ${B}_t\setminus{A}_t$ to check reduces by at least $1$ in each iteration. Therefore, it takes at most $O\big((|{V}|+|{V}|-1+\dots+1)M\big)=O(|{V}|^2M)$ time to find ${A}^\ast$ and ${B}^\ast$. After the reduction of the search space, there are $k_1=|{B}^\ast\setminus{A}^\ast|$ nodes to be further examined. For the Greedy algorithm, it checks $k_1-i+1$ nodes in the $i$th iteration. Thus, it takes at most $O(k_1^2M)$ time, which means the total time complexity of the Greedy algorithm is $O\big((|{V}|^2+k_1^2)M\big)$. Each iteration of the ModMod algorithm has a time complexity of $O(k_1M)$. Let $k_{2}$ denote the total number of iterations used for the ModMod algorithm. Then, the total time complexity of the ModMod algorithm is $O\big((|{V}|^2+k_1k_2)M\big)$.

\textbf{Diffusion Models:}
Our analysis and algorithms are general frameworks that can be adapted to any diffusion models which are submodular, such as the Independent Cascade and Linear Threshold models, the triggering model \cite{Kempe_maxInfluence_2003}, the continuous-time models \cite{Chen_time_2012,Du_CTIC_2013}, and the topic-aware models \cite{Barbieri_TAI_2012,Chen_TAI_2015}.

\section{Performance Analysis}\label{sec:analysis}
The challenges to evaluate the quality of the seed set constructed for the profit maximization problem are two-fold. First, optimizing the difference between two submodular functions is multiplicative inapproximability unless P=NP. Thus, it is difficult to measure the \textit{gap} between the seed set obtained and an optimal seed set. Second, the random processes of many diffusion models are analytically intractable. For example, computing the exact influence spread under the IC diffusion model is \#P-hard \cite{Chen_MIA_2010}. Thus, the benefit brought and the cost incurred by a seed set can only be estimated via some sampling approaches \cite{Kempe_maxInfluence_2003,Borgs_RIS_2014}. As a result, the \textit{sampling error} also affects the quality measurement of the seed set.

We propose techniques to analyze the aforementioned gap and sampling error, which enable us to evaluate the approximation guarantee of the seed set obtained by any algorithm on any given instance of the profit maximization problem. Specifically, let ${S}^o$ be the seed set constructed for a problem instance. We develop an upper bound $\mu$ on the maximum achievable profit for the problem instance to characterize the gap between the real profit value $\phi({S}^o)$ and the maximum achievable profit. Note that both $\phi({S}^o)$ and $\mu$ are to be estimated by sampling. Let $\tilde{\phi}({S}^o)$ and $\tilde{\mu}$ be their estimated values. We further study the sampling errors to bound the difference between $\tilde{\phi}({S}^o)$ and $\phi({S}^o)$ and the difference between $\tilde{\mu}$ and $\mu$. In this way, we can obtain an approximation guarantee of ${S}^o$ using the estimated values $\tilde{\phi}({S}^o)$ and $\tilde{\mu}$.

\subsection{Upper Bound of Maximum Achievable Profit}\label{subsec:online_bounds}
To derive our bounds on the maximum achievable profit, we first introduce two modular bounds for submodular functions.

\textbf{Modular Upper Bounds:}
For any submodular set function $f(\cdot)$, we have the following two modular upper bounds $m_{X}^1$ and $m_{X}^2$ that are tight at a given set ${X}$ \cite{Iyer_submodularDiff_2012}:
\begin{align}
&m_{X}^1({Y})\triangleq f({X})-\!\!\!\sum_{v\in{X}\setminus{Y}}\!\!\!f(v\mid{V}\setminus\{v\})+\!\!\!\sum_{v\in{Y}\setminus{X}}\!\!\!f(v\mid{X}),\label{eq:mod_o1}\\
&m_{X}^2({Y})\triangleq f({X})-\!\!\!\sum_{v\in{X}\setminus{Y}}\!\!\!f(v\mid{X}\setminus\{v\})+\!\!\!\sum_{v\in{Y}\setminus{X}}\!\!\!f(v\mid\emptyset).\label{eq:mod_o2}
\end{align}
In the previous section, we have reduced the search space so that only the sets belonging to $[{A}^\ast,{B}^\ast]$ need to be considered for profit maximization. As a result, for any ${A}^\ast\subseteq{X},{Y}\subseteq{B}^\ast$, the above two upper bounds can be improved to:
\allowdisplaybreaks[4]
\begin{align}
&\!\!\!m_{X}^3({Y})\triangleq f({X})-\!\!\!\sum_{v\in{X}\setminus{Y}}\!\!\!f(v\mid{B}^\ast\setminus\{v\})+\!\!\!\sum_{v\in{Y}\setminus{X}}\!\!\!f(v\mid{X}),\label{eq:mod1}\\
&\!\!\!m_{X}^4({Y})\triangleq f({X})-\!\!\!\sum_{v\in{X}\setminus{Y}}\!\!\!f(v\mid{X}\setminus\{v\})+\!\!\!\sum_{v\in{Y}\setminus{X}}\!\!\!f(v\mid{A}^\ast).\label{eq:mod2}
\end{align}
It is easy to show that the bounds $m_{X}^3({Y})$ and $m_{X}^4({Y})$ remain tight at ${X}$, i.e., $m_{X}^3({X})=m_{X}^4({X})=f({X})$, and they are tighter than $m_{X}^1({Y})$ and $m_{X}^2({Y})$ at other sets, i.e., $m_{X}^1({Y})\geq m_{X}^3({Y})\geq f({Y})$ and $m_{X}^2({Y})\geq m_{X}^4({Y})\geq f({Y})$ for any ${A}^\ast\subseteq{Y}\subseteq{B}^\ast$.

\textbf{Modular Lower Bounds:} 
For any submodular set function $f(\cdot)$, a modular lower bound $h_{X}$ that is tight at a given set ${X}$ can be obtained as follows \cite{Fujishige_submodularOpt_2005}. Let $\pi$ be any permutation of ${V}$ that places all the nodes in ${X}$ before the nodes in ${V}\setminus{X}$. Let ${S}^\pi_i=\{\pi(1),\pi(2),\cdots,\pi(i)\}$ be a chain formed by the permutation, where ${S}^\pi_0=\emptyset$ and ${S}^\pi_{|{X}|}={X}$. Define  
\begin{equation}\label{eq:marginal_lower}
h_{X}^\pi(\pi(i)) = f({S}^\pi_i)-f({S}^\pi_{i-1}).
\end{equation}
Then, $h_{X}^\pi({Y})=\sum_{v\in{Y}} h_{X}^\pi(v)$ is a lower bound of $f({Y})$, which is tight at ${X}$, i.e., $h_{X}^\pi({Y})\leq f({Y})$ for any ${Y}\subseteq{V}$ and $h_{X}^\pi({X})=f({X})$. After the search space is reduced to $[{A}^\ast,{B}^\ast]$, we restrict $\pi$ to any permutation of ${V}$ in the order of ${A}^\ast$, ${X}\setminus{A}^\ast$ and ${B}^\ast\setminus{X}$.

\textbf{Upper Bounds on Maximum Achievable Profit:} 
Based on the above bounds, we can derive two series of upper bounds on the maximum value of the profit function $\phi(\cdot)$ as follows. For any set ${X}$ where ${A}^\ast\subseteq{X}\subseteq{B}^\ast$, we define 
\begin{equation}\label{eq:ubounds}
\mu_i({X})\triangleq\!\!\max_{{A}^\ast\subseteq{Y}\subseteq{B}^\ast}\!\!m_{X}^i({Y};\beta)-h_{X}^{\pi}({Y};\gamma),
\end{equation}
where $m_{X}^i({Y};\beta)$ ($i=3,4$) denotes the modular upper bound on the benefit function $\beta$ and $h_{X}^{\pi}({Y};\gamma)$ denotes the modular lower bound on the cost function $\gamma$ respectively. For briefness, we shall use $\mu({X})$ to refer to either bound with $i=3$ or $4$ in the rest of the paper. 
\begin{theorem}
	\label{theorem:modularGlobalBound} 
	For any set ${X}$ where ${A}^\ast\subseteq{X}\subseteq{B}^\ast$,
	\begin{equation}
	\mu({X})\geq\max_{{S}\subseteq{V}}\phi({S}).
	\end{equation}
\end{theorem}

The upper bounds established above can be computed very fast since $m_{X}^i({Y};\beta)$ and $h_{X}^{\pi}({Y};\gamma)$ are both modular functions with respect to ${Y}$. It is much easier to find the maximum value for a modular function than that for a submodular function. We can obtain upper bounds by arbitrarily choosing the set ${X}$ in $\mu({X})$. Given a seed set solution ${S}^o$ obtained by any algorithm, we simply choose ${X}={S}^o$. Then, the approximation guarantee of ${S}^o$ can be estimated by $\phi({S}^o)/\mu({S}^o)$.

\subsection{Sampling Error}\label{subsec:sampling_error}
To estimate $\phi({S}^o)$ and $\mu({S}^o)$, we make use of a state-of-the-art technique called reverse influence sampling \cite{Borgs_RIS_2014,Nguyen_BCT_2016}.

\begin{definition}[RR Set for Weighted Graph]\label{definition:RR}
	A random reverse reachable (RR) set ${R}$ for a weighted graph ${G}$ is generated by (1) first selecting a random node $v\in{V}$ with a probability distribution $p(\cdot)$ proportional to the node weights, (2) then sampling a graph $g$ randomly from ${G}$ according to the diffusion model, (3) finally taking the set of nodes in $g$ that can reach $v$ as ${R}$.
\end{definition}

Taking our benefit metric as an example, the probability for choosing a node $v$ in a random RR set is given by $p(v)=b(v)/\Upsilon_b$, where $\Upsilon_b=b(V)=\sum_{v\in{V}}b(v)$ is the total benefit weight of all nodes. For a seed set ${S}$, the relation between its benefit and a random RR set ${R}$ is $\beta({S})=\Upsilon_b\cdot\Pr[{S}\cap{R}\neq \emptyset]$, where $\Pr[{S}\cap{R}\neq \emptyset]$ is the probability that ${R}$ contains at least one node in ${S}$ \cite{Nguyen_BCT_2016}. Similarly, we can estimate the diffusion cost $\gamma({S})$ by choosing each node in a random RR set with a probability proportional to the cost of the node, i.e., $p(v)=c(v)/\Upsilon_c$, where $\Upsilon_c=c(V)=\sum_{v\in{V}}c(v)$ is the total cost weight of all nodes. Therefore, we generate two groups of RR sets to estimate the benefit and cost respectively for a seed set.

Suppose that we generate a total of $\theta_\beta$ random RR sets to estimate the benefit brought by a seed set ${S}$. An RR set is said to be covered by ${S}$ if it contains at least one node in ${S}$. Let $\Lambda_\beta({S})$ denote the number of RR sets covered by ${S}$ among the $\theta_\beta$ random RR sets. Then, the benefit brought by ${S}$ can be estimated as $\Lambda_\beta({S})\cdot \Upsilon_b/\theta_\beta$. To analyze the sampling error, we make use of the Chernoff-Hoeffding Theorem \cite{Dagum_MCestimation_2000}.
\begin{lemma}[Chernoff-Hoeffding Theorem \cite{Dagum_MCestimation_2000}]\label{lemma:Chernoff}
	Let $Z_1,Z_2,\dots,Z_\theta$ denote random variables that are independently and identically distributed according to $Z$ in the interval $[0,1]$ with mean $\mathbb{E}[Z]$. For any fixed $\theta>0$ and $\varepsilon>0$,
	\begin{align*}
	&\Pr\left[\sum_{i=1}^{\theta}Z_i - \theta\cdot\mathbb{E}[Z]\geq \varepsilon\right]\leq \exp\left(-\frac{\varepsilon^2}{4(e-2)\theta\mathbb{E}[Z]}\right),\\
	&\Pr\left[\sum_{i=1}^{\theta}Z_i - \theta\cdot\mathbb{E}[Z]\leq -\varepsilon\right]\leq \exp\left(-\frac{\varepsilon^2}{4(e-2)\theta\mathbb{E}[Z]}\right).
	\end{align*}
\end{lemma}

Based on Lemma~\ref{lemma:Chernoff}, we can establish the following relation between $\Lambda_\beta({S})$ and the real benefit $\beta({S})$.

\begin{theorem}\label{theorem:sample_error}
	For $\theta_\beta$ random RR sets that are independent of ${S}$ and any $\delta\in(0,1)$, we have
	\begin{align*}
	&\Pr\left[\beta({S})\geq \left(\sqrt{\Lambda_\beta({S})+0.25a} - 0.5\sqrt{a}\right)^2\!\!\cdot \frac{\Upsilon_b}{\theta_\beta}\right]\geq 1-\frac{\delta}{2},\\
	&\Pr\left[\beta({S})\leq \left(\sqrt{\Lambda_\beta({S})+0.25a}+0.5\sqrt{a}\right)^2\!\!\cdot \frac{\Upsilon_b}{\theta_\beta}\right]\geq 1-\frac{\delta}{2},
	\end{align*}
	where $a=4(e-2)\ln(2/\delta)$.
\end{theorem}

According to Theorem~\ref{theorem:sample_error}, given $\Lambda_\beta({S})$, we can define a lower bound and an upper bound of $\beta({S})$ with a probability at least $1-\delta/2$ as
\begin{equation}
\begin{cases}
\beta_l({S})\triangleq\left(\sqrt{\Lambda_\beta({S})+0.25a} - 0.5\sqrt{a}\right)^2\cdot \Upsilon_b/\theta_\beta,\\
\beta_u({S})\triangleq\left(\sqrt{\Lambda_\beta({S})+0.25a}+0.5\sqrt{a}\right)^2\cdot \Upsilon_b/\theta_\beta.
\end{cases}
\end{equation}

The above analysis can also be applied to the estimation of the cost. Let $\theta_\gamma$ denote the number of random RR sets generated to estimate the cost incurred by a seed set ${S}$. Let $\Lambda_\gamma({S})$ denote the number of RR sets covered by ${S}$ among the $\theta_\gamma$ random RR sets. Then, the cost incurred by ${S}$ can be estimated as $\Lambda_\gamma({S})\cdot \Upsilon_c/\theta_\gamma$. We can similarly define a lower bound $\gamma_l({S})$ and an upper bound $\gamma_u({S})$ on the real cost $\gamma({S})$ with a probability at least $1-\delta/2$. Then, given a seed set solution ${S}^o$ obtained by any algorithm, we can derive a lower bound $\beta_l({S}^o)-\gamma_u({S}^o)$ on its real profit $\phi({S}^o)$ such that $\Pr[\phi({S}^o)\geq \beta_{l}({S}^o)-\gamma_{u}({S}^o)]\geq \Pr\big[\big(\beta({S}^o)\geq \beta_{l}({S}^o)\big)\wedge\big(\gamma({S}^o)\leq\gamma_{u}({S}^o)\big)\big]=1-\Pr\big[\big(\beta({S}^o)< \beta_{l}({S}^o)\big)\vee\big(\gamma({S}^o)>\gamma_{u}({S}^o)\big)\big]\geq1-(\delta/2+\delta/2)>1-\delta$.

The difficulty in analyzing the sampling error for the upper bound $\mu({S}^o)$ of the maximum achievable profit lies in that we can never find a specific seed set to achieve the upper bound. From Theorem~\ref{theorem:sample_error}, we know that the profit of any seed set ${S}$ is bounded above by $\beta_u({S})-\gamma_l({S})$ with a high probability. By definition, $\beta_u({S})-\gamma_l({S})$ is a function of $\Lambda_\beta({S})$ and $\Lambda_\gamma({S})$. Let $\tilde{\phi}({S})$ be the estimated profit of ${S}$ on the RR sets generated, which is defined by $\tilde{\phi}({S})=\Lambda_\beta({S})\cdot \Upsilon_b/\theta_\beta-\Lambda_\gamma({S})\cdot \Upsilon_c/\theta_\gamma$. Then, $\beta_u({S})-\gamma_l({S})$ can be represented as a function $\eta$ of $\Lambda_\beta({S})$ and $\tilde{\phi}({S})$, i.e., $\eta\big(\Lambda_\beta({S}),\tilde{\phi}({S})\big)$. We can prove that $\eta\big(\Lambda_\beta({S}),\tilde{\phi}({S})\big)$ is increasing with both $\Lambda_\beta({S})$ and $\tilde{\phi}({S})$. Naturally, $\Lambda_\beta({S})$ is bounded above by $\theta_\beta$. On the other hand, thanks to the submodularity of the set coverage used to estimate the benefit and cost metrics in the reverse influence sampling approach, we can easily get the estimated upper bound $\tilde{\mu}({S}^o)$ on the maximum achievable profit from the RR sets generated according to the analysis in Section~\ref{subsec:online_bounds}. As a result, for any seed set ${S}$, the upper bound $\eta\big(\Lambda_\beta({S}),\tilde{\phi}({S})\big)$ of its profit is bounded above by $\eta\big(\theta_\beta,\tilde{\mu}({S}^o)\big)$. In this way, we can obtain the sampling error for the upper bound $\mu({S}^o)$.
\begin{theorem}\label{theorem:ds_optimal_upper_bound}
	For any $\delta\in(0,1)$, we have 
	\begin{equation}\label{eq:ds_upper_bound_optimal}
	\Pr\left[\max_{{S}\subseteq{V}}\phi({S})\leq \tilde{\mu}({S}^o)+\varepsilon\big(\tilde{\mu}({S}^o)\big)\right]\geq 1-\delta,
	\end{equation}
	where $\varepsilon\big(\tilde{\mu}({S}^o)\big)$ is the sampling error for $\tilde{\mu}({S}^o)$ such that $\varepsilon\big(\tilde{\mu}({S}^o)\big)=\rho_\gamma\sqrt{a\Big(\big(\rho_\beta\theta_\beta-\tilde{\mu}({S}^o)\big)/\rho_\gamma+0.25a\Big)}+0.5a(\rho_\beta-\rho_\gamma)+\rho_\beta\sqrt{a(\theta_\beta+0.25a)}$, and $\rho_\beta=\Upsilon_b/\theta_\beta$ and $\rho_\gamma=\Upsilon_c/\theta_\gamma$.
\end{theorem}

By Theorems~\ref{theorem:sample_error} and \ref{theorem:ds_optimal_upper_bound}, we have the approximation guarantee that
\begin{equation}\label{eq:guarantee}
\frac{\phi({S}^o)}{\max_{{S}\subseteq{V}}\phi({S})}\geq\frac{\beta_{l}({S}^o)-\gamma_{u}({S}^o)}{\tilde{\mu}({S}^o)+\varepsilon\big(\tilde{\mu}({S}^o)\big)}
\end{equation}
with a probability at least $1-2\delta$.

\subsection{Reduce Sampling Error via Normalization}\label{subsec:normalization}
As discussed in Section \ref{subsec:problem}, we can normalize the benefit and cost weights by $\bar{b}(v)$ and $\bar{c}(v)$ for every node $v\in{V}$. Intuitively, the normalization can avoid unnecessary samples conducted by the weights $\sum_{v\in{V}}\min\{b(v),c(v)\}$ for the estimations of both the benefit and cost. Therefore, the normalization can reduce the sampling error by increasing the number of useful samples. 

\begin{theorem}\label{theorem:normalization}
	For any seed set ${S}$ and a fixed number of samples, let $\varepsilon_\phi$ be the sampling error limit that can provide a probability guarantee of $1-\delta$, i.e., $\Pr[-\varepsilon_\phi\leq\tilde{\phi}({S})-\phi({S})\leq\varepsilon_\phi]\geq 1-\delta$, and let $\bar{\varepsilon}_\phi$ be the sampling error limit under the normalized form. We have $\bar{\varepsilon}_\phi\leq \varepsilon_\phi$.
\end{theorem}

Theorem~\ref{theorem:normalization} indicates that the normalization can improve the solution quality which shall be demonstrated in the experiments.

\section{Evaluation}\label{sec:evaluation}

\begin{figure*}[!t]
	\centering
	\includegraphics[width=1.0\linewidth]{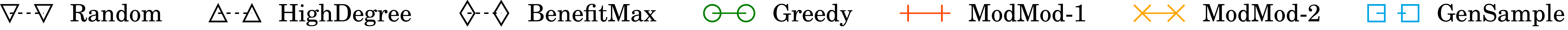}\vspace{-0.15in}\\
	\begin{minipage}{.49\textwidth}
		\capstart
		\subfloat[Google+]{\includegraphics[width=0.48\linewidth]{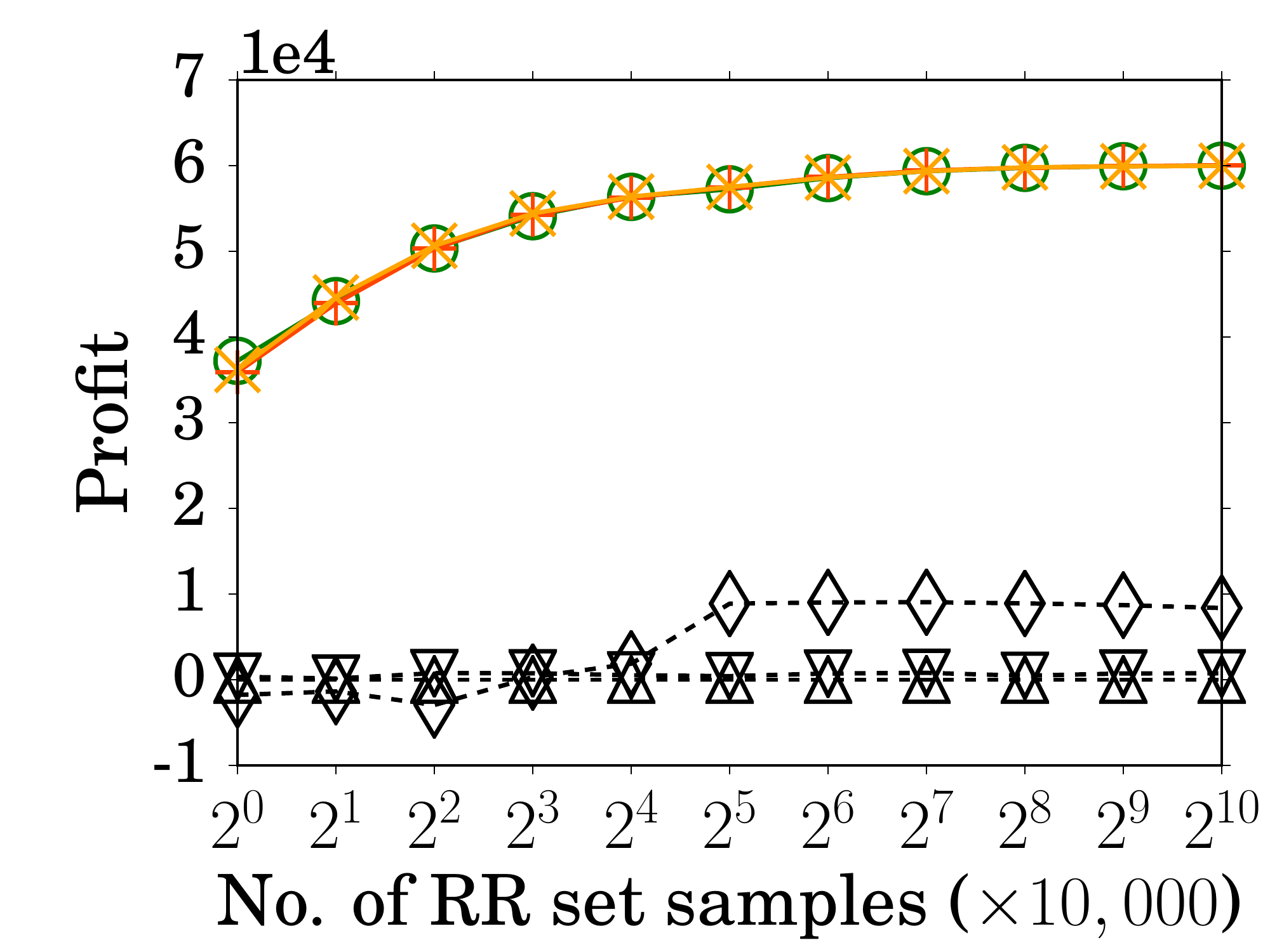}\label{subfig:gplus_profit_norm}}\hfil
		\subfloat[LiveJournal]{\includegraphics[width=0.48\linewidth]{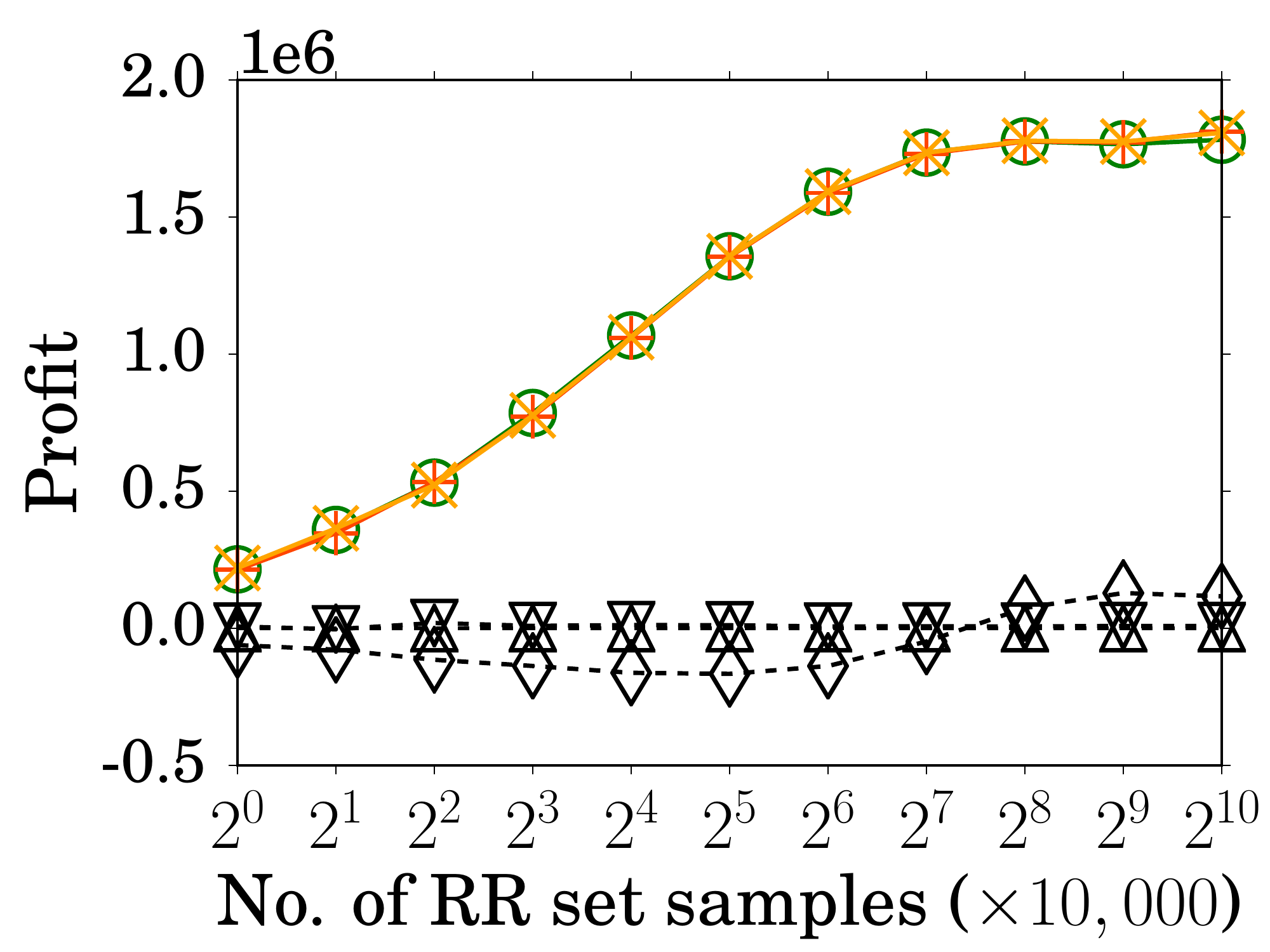}\label{subfig:liveJournal_profit_norm}}\hfill
		\caption{Profits produced by different algorithms.}\label{fig:profit_norm}
	\end{minipage}
	\begin{minipage}{.49\textwidth}
		\capstart
		\subfloat[Google+]{\includegraphics[width=0.48\linewidth]{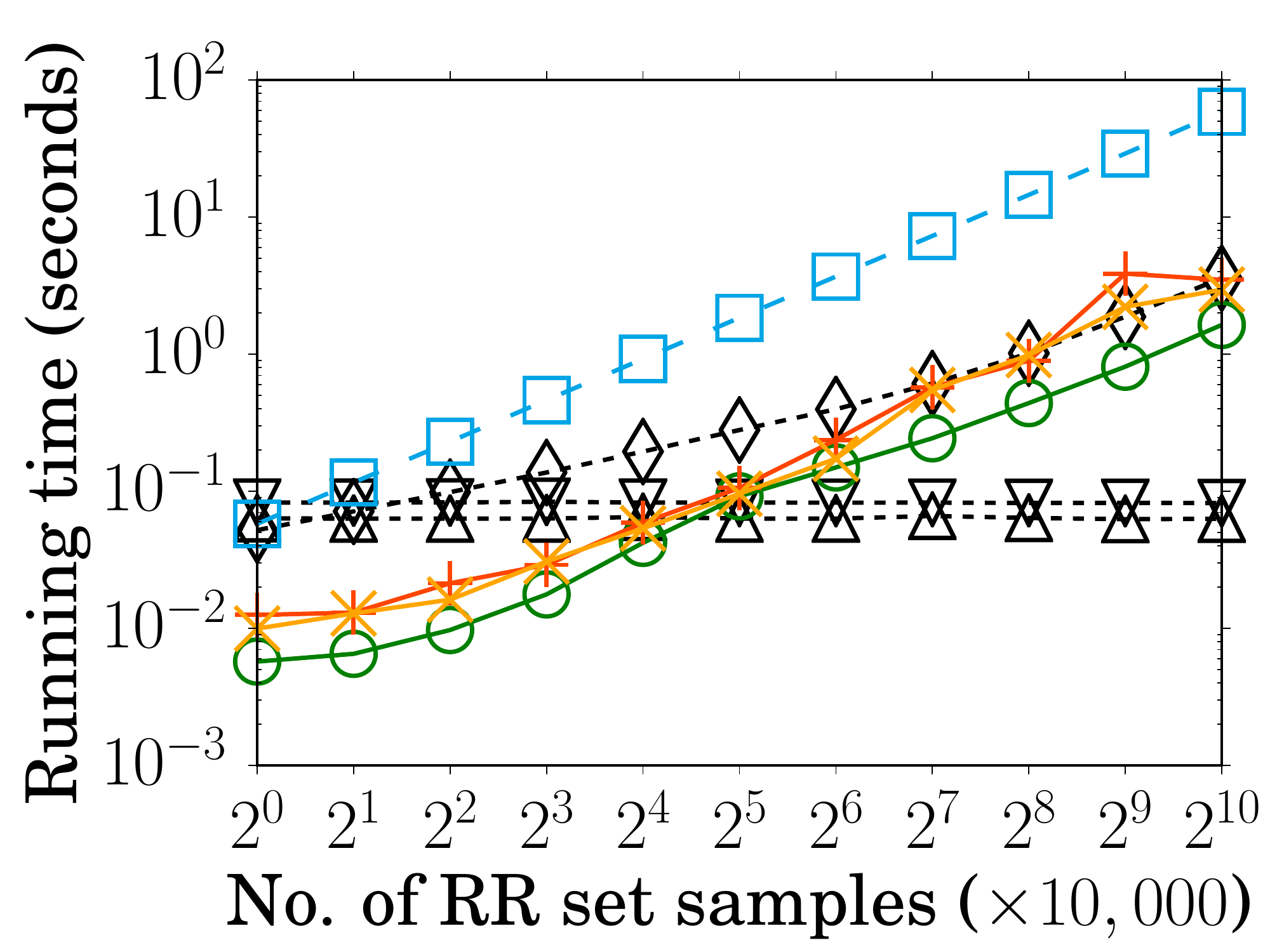}\label{subfig:gplus_time_norm}}\hfil
		\subfloat[LiveJournal]{\includegraphics[width=0.48\linewidth]{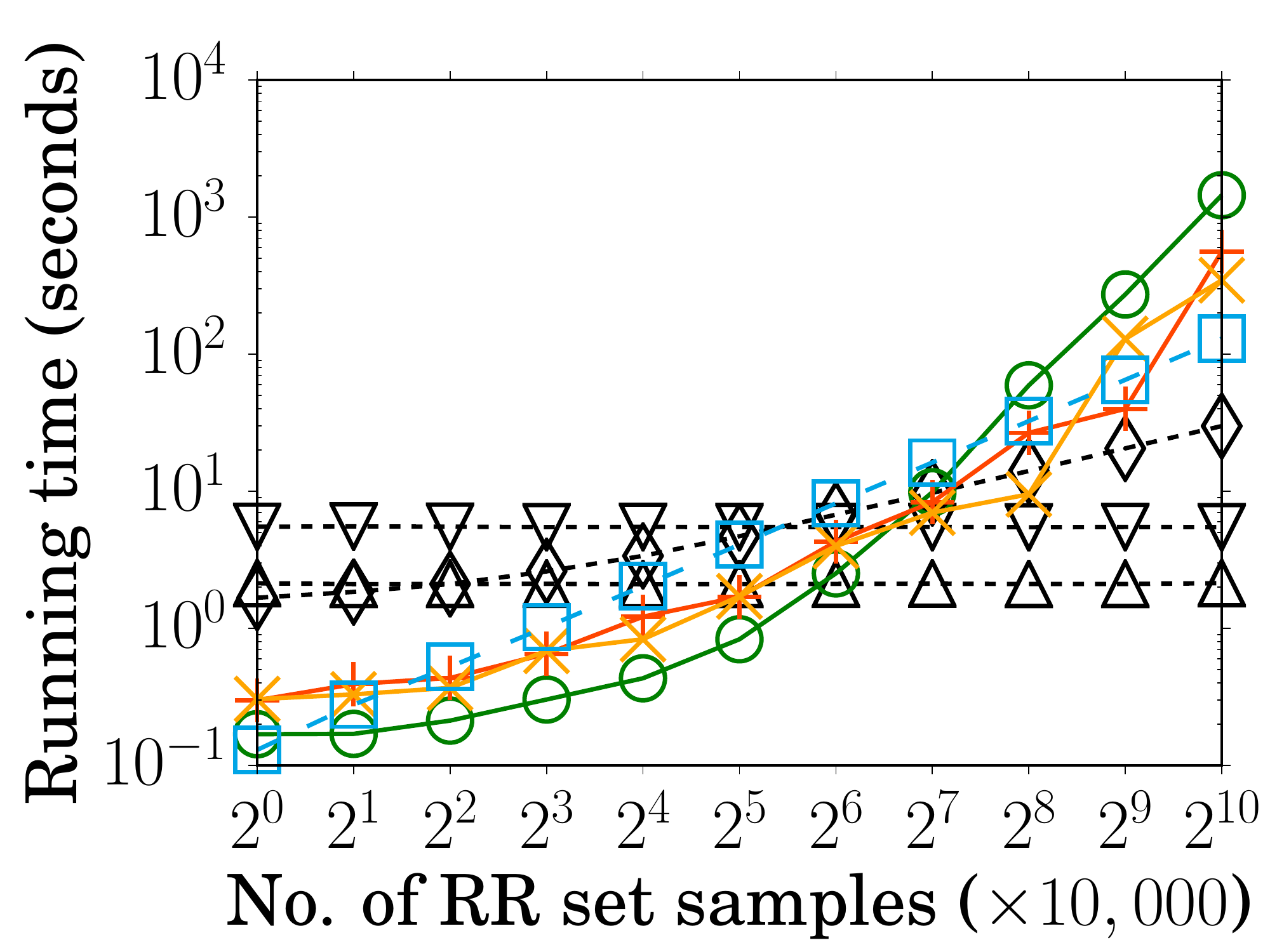}\label{subfig:liveJournal_time_norm}}\hfill
		\caption{Running times of different algorithms.}\label{fig:time_norm}
	\end{minipage}
	\vspace{-0.05in}
\end{figure*}

\begin{figure*}[!t]
	\capstart
	\centering
	\includegraphics[width=1.0\linewidth]{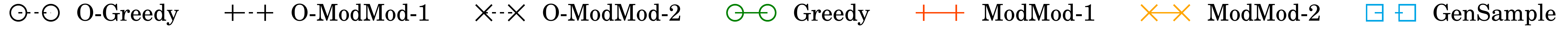}\vspace{-0.15in}\\
	\subfloat[Google+]{\includegraphics[width=0.235\linewidth]{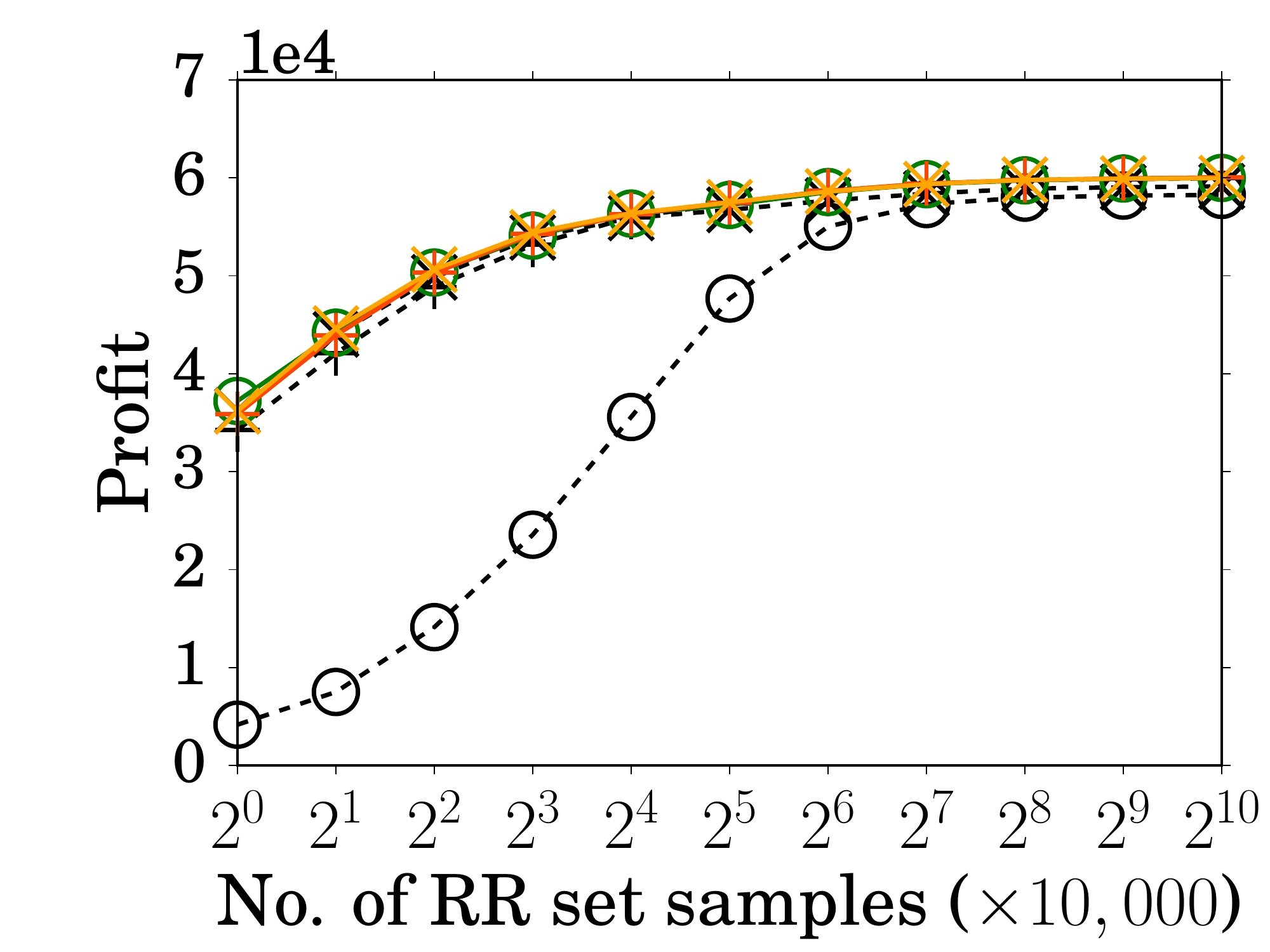}\label{subfig:gplus_profit_prune}}\hfill
	\subfloat[LiveJournal]{\includegraphics[width=0.235\linewidth]{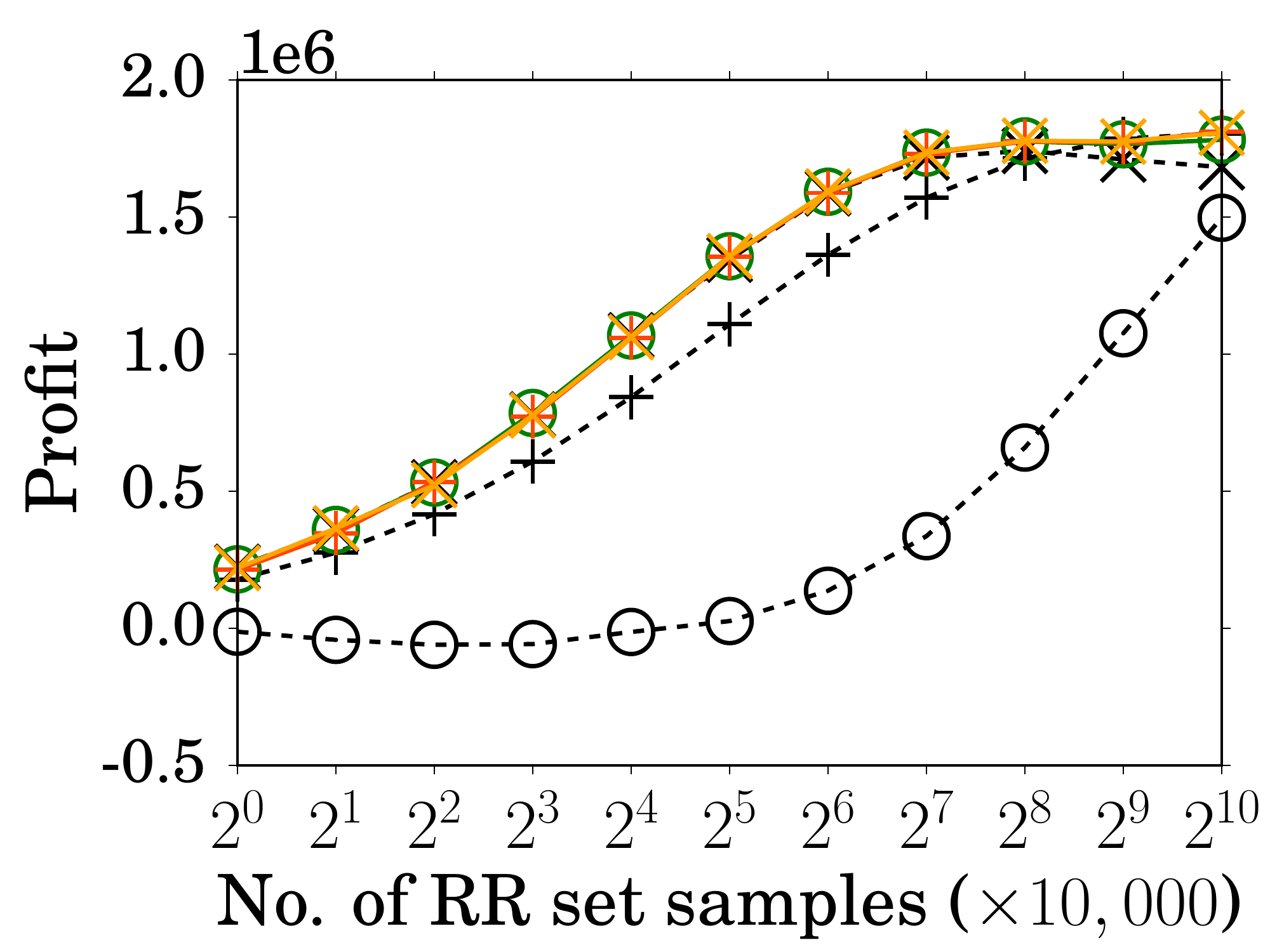}\label{subfig:liveJournal_profit_prune}}\vspace{0.15in}\hfill
	\subfloat[Google+]{\includegraphics[width=0.235\linewidth]{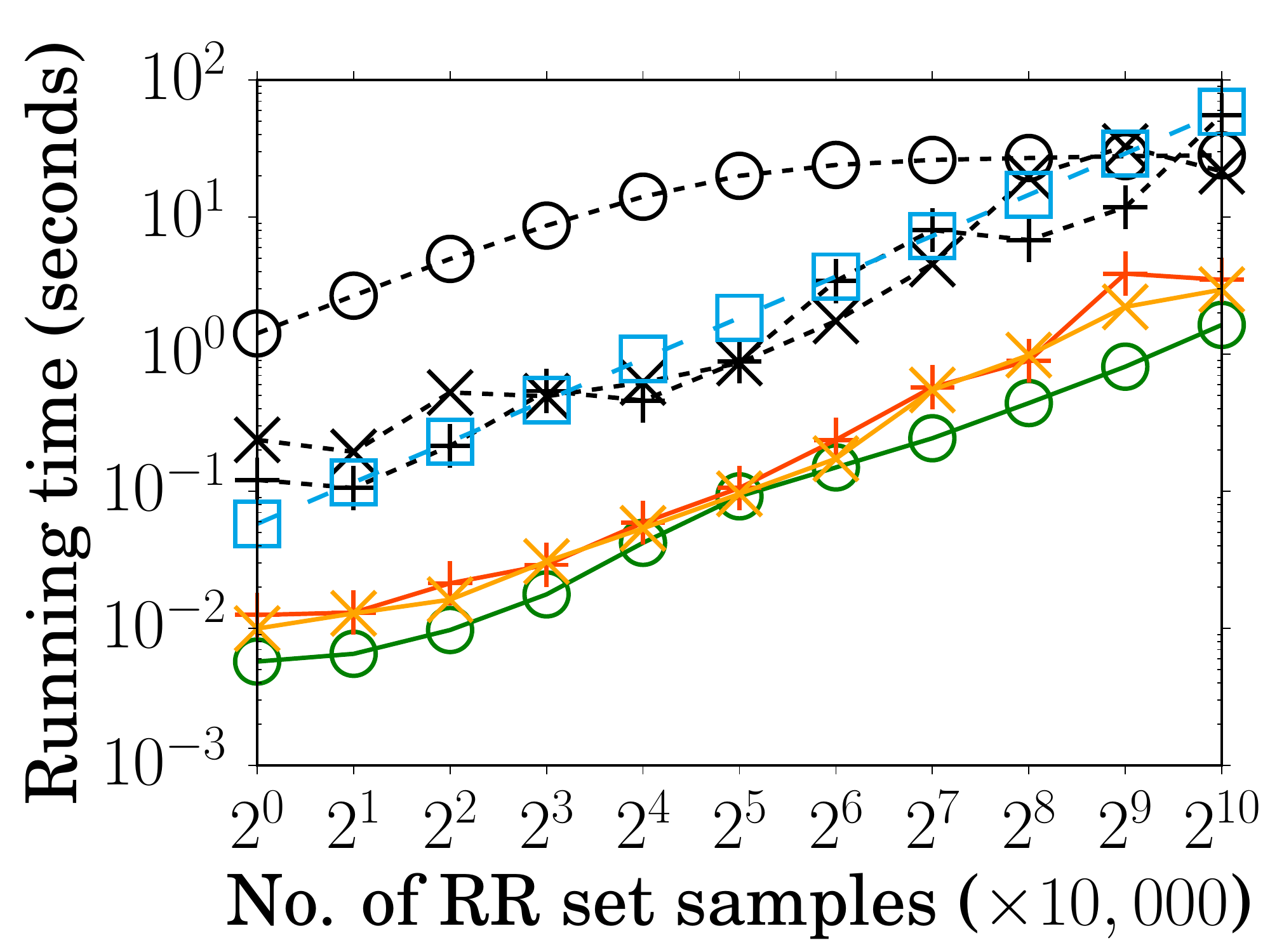}\label{subfig:gplus_time_prune}}\hfill
	\subfloat[LiveJournal]{\includegraphics[width=0.235\linewidth]{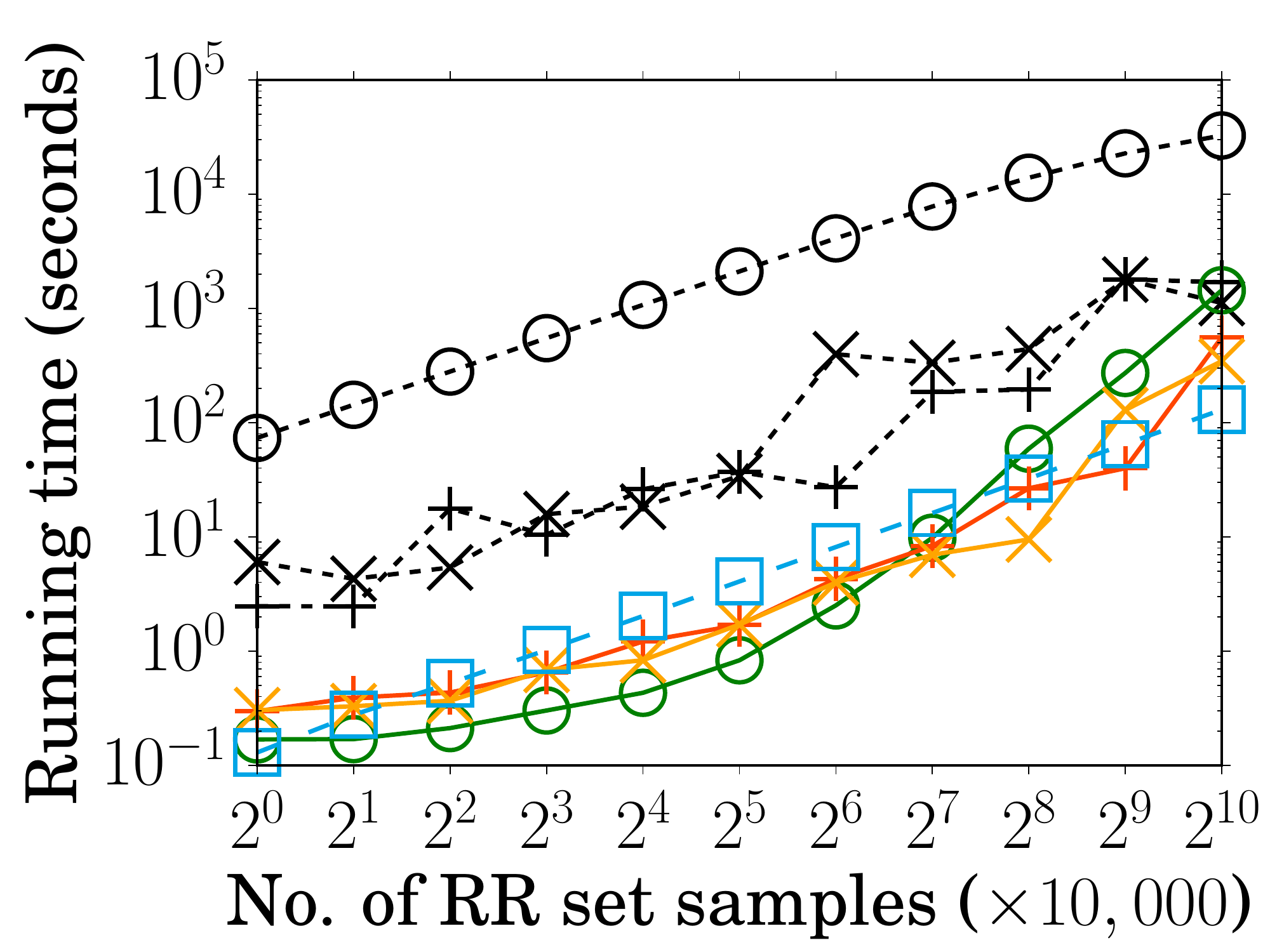}\label{subfig:liveJournal_time_prune}}\hfill
	\vspace{-0.15in}
	\caption{Impact of iterative pruning technique (algorithms with prefix ``O-'' do not use pruning technique).}\label{fig:profit_prune}
	\vspace{-0.15in}
\end{figure*}

\subsection{Experimental Setup}\label{subsec:setup}
\textbf{Datasets.} We use several real social networks available at \cite{Leskovec_SNAP_2014} to evaluate our proposed techniques. Due to space limitations, we report here the results for two representative datasets, Google+ ($108$K nodes, $14$M edges), and LiveJournal ($5$M nodes, $69$M edges).

\textbf{Algorithms.} Recall that the ModMod algorithm needs a modular lower bound of the benefit function $\beta(\cdot)$ and a modular upper bound of the cost function $\gamma(\cdot)$. In Section~\ref{subsec:online_bounds}, we have presented one such lower bound and two such upper bounds. We use ModMod-1 to refer to that using the upper bound $m_{X}^3({Y})$ defined in Eq. \eqref{eq:mod1} and use ModMod-2 to refer to that using the upper bound $m_{X}^4({Y})$ defined in Eq. \eqref{eq:mod2}. We compare our two-phase methods with the following baselines.

\begin{itemize}
	\item \textit{Random}: It randomly selects $k$ nodes. We run the algorithm $10$ times and take their average as the expected profit. 
	\item \textit{HighDegree}: It selects $k$ nodes with the highest degrees.
	\item \textit{BenefitMax}: It makes use of the reverse influence sampling technique to find the top-$k$ influential nodes for influence/benefit maximization \cite{Borgs_RIS_2014,Nguyen_BCT_2016,Nguyen_DSSA_2016,Tang_IMM_2015,Tang_reverse_2014}.
\end{itemize}
The above baselines are executed on the entire social networks without applying any pruning technique. To explore different seed numbers, in each baseline, we iterate through $k=\frac{|{V}|}{2^i}$ for $i=0,1,\dots,10$ (where $|{V}|$ is the network size) and choose the $k$ value producing the largest profit.

\textbf{Parameter Settings.} By default, we use the IC diffusion model (as described in Example~\ref{example:pruning} of Section~\ref{subsection:pruning}), a uniform benefit distribution (where every node has a unit benefit to model the commission paid by the advertiser for each user activated), and a degree-proportional cost distribution (where the cost of each node is set proportional to its out-degree to emulate the diffusion cost for each activated user to push the product advertisement to all of his neighbors). In the IC model, we set the propagation probability $p_{u,v}$ of each edge $\langle{u,v}\rangle$ to the reciprocal of $v$'s in-degree (the number of $v$'s inverse neighbors) as widely adopted by other studies \cite{Chen_MIA_2010,Jung_IRIE_2012,Nguyen_BCT_2016,Nguyen_DSSA_2016,Tang_IMM_2015,Tang_reverse_2014}. 

By default, we normalize the benefit and cost weights as described in Section~\ref{subsec:problem}. We use a scale factor $r$ to control the ratio between the total cost and total benefit of all nodes. A higher $r$ implies a higher cost of influence propagation relative to the benefit of influence spread. The default value of $r$ is set to $1$. We have tested a wide range of $r$ values and observed similar performance trends. In executing our two-phase methods and the baseline BenefitMax algorithm, we vary the number of RR sets generated to study the impact of benefit and cost estimations. To evaluate the profits of the seed sets returned by different algorithms, we generate a group of validation RR sets to keep the estimation errors within $1\%$ with a high probability at least $1-10^{-6}$ according to Theorem~\ref{theorem:sample_error}.

\subsection{Profits Produced by Different Algorithms}\label{subsec:result}

\figurename~\ref{fig:profit_norm} shows the profits produced by different algorithms. Comparing the seed selection algorithms, our heuristic algorithms are more effective in optimizing the profit than the three baseline algorithms (Random, HighDegree and BenefitMax) on the datasets tested. This suggests that improving the influence spread or benefit alone is not effective for maximizing the profit. Our three heuristic algorithms perform quite close in terms of the profit produced. It can also be seen that with increasing number of RR sets generated for profit estimation in the seed selection process, the solution quality of our heuristics is improved due to lower estimation errors.

\subsection{Running Times of Different Algorithms}\label{subsec:result_time}
\figurename~\ref{fig:time_norm} shows the running times of different algorithms. The algorithms are all implemented in C++ and the experiments are carried out on a machine with an Intel Xeon E5-1650 3.2GHz CPU and 16GB memory. The time spent for generating RR sets is common to all our heuristics as well as BenefitMax. We plot it as a separate curve and exclude it from the running times of all the algorithms in \figurename~\ref{fig:time_norm}. It can be seen that generating the RR sets takes significant time. The Random and HighDegree algorithms do not need benefit and cost estimations. Thus, their running times are independent of the number of RR sets. The other four methods have running times increasing almost linearly with the number of RR sets (note that both axes are in logscale). In most cases, our heuristic methods complete execution within $100$ seconds even for the LiveJournal dataset with millions of nodes. This shows the efficiency of our two-phase algorithms.

\subsection{Iterative Pruning Technique}\label{subsec:result_prune}
\begin{table}[!t]
	\capstart
	\centering
	\caption{Search space reduction.}
	\label{table:pruning}
	\vspace{-0.15in}
	\setlength{\tabcolsep}{0.4em} 
	\renewcommand{\arraystretch}{1.2}
	\subfloat[With Normalization]{
		\label{subtable:prune_norm}
		\begin{tabular}{c||cccc|c}
			Dataset & $|{V}|$ & $|{A}^\ast|$ & $|{B}^\ast|$ & $|{B}^\ast\setminus{A}^\ast|$ & Reduction\\ \hline
			Google+ & $108$K & $80.6$K & $82.4$K & $1.8$K & $98.3\%$ \\
			LiveJournal & $5$M & $2.5$M & $3.3$M & $0.8$M & $83.7\%$ \\
		\end{tabular}
	}\vfill
	\subfloat[Without Normalization]{
		\label{subtable:prune_non-norm}
		\begin{tabular}{c||cccc|c}
			Dataset & $|{V}|$ &  $|{A}^\ast|$ & $|{B}^\ast|$ & $|{B}^\ast\setminus{A}^\ast|$ & Reduction\\ \hline
			Google+ & $108$K & $62.2$K & $97.1$K & $34.9$K & $67.6\%$ \\
			LiveJournal & $5$M & $1.3$M & $4.3$M & $3.0$M & $38.3\%$ \\
		\end{tabular}
	}
	\vspace{-0.25in}
\end{table}

Table~\ref{table:pruning} shows the amount of search space reduction by the iterative pruning technique proposed in Section~\ref{subsection:pruning}. These results are for the experiments with $2^{10}\times 10,000$ RR sets generated. As can be seen, the pruning technique substantially reduces the number of nodes that need to be considered for seed selection. Specifically, the search space is reduced by $98.3\%$ and $83.7\%$ of the original network size for the Google+ and LiveJournal datasets respectively when the normalization is applied. On the other hand, the reduction is much less when the normalization is not applied ($67.6\%$ and $38.3\%$ for the Google+ and LiveJournal datasets respectively). This observation confirms Theorem~\ref{theorem:normalization_prune}. 

\figurename~\ref{fig:profit_prune} shows the impact of the search space reduction on the profit produced and the running time of our heuristics, where O-X refers to heuristic X without the search space reduction. As claimed in Corollary~\ref{corollary:pruneImprovement}, the pruning technique can help improve all the heuristic algorithms in terms of the profit produced. This is confirmed by the experimental results. For the LiveJournal dataset (\figurename~\ref{subfig:liveJournal_profit_prune}), the Greedy algorithm even produces negative profit without pruning the search space first, and the pruning technique can bring improvements up to $24.1\%$ and $14.3\%$ for the ModMod-1 and ModMod-2 algorithms respectively. Furthermore, the pruning technique can also help reduce the running time considerably. By pruning the search space first, our heuristic algorithms can run up to $3$ orders faster. These observations demonstrate the effectiveness of our pruning technique.

\subsection{Guarantee of Solution Quality}\label{subsec:result_bounds}
\begin{table}[!t]
	\vspace{-0.02in}
	\capstart
	\centering
	\caption{Approximation guarantee of seed sets obtained.}
	\label{table:bounds}
	\setlength{\tabcolsep}{0.4em} 
	\renewcommand{\arraystretch}{1.2}
	\begin{tabular}{c||ccc}
		Dataset & Greedy & ModMod-1 & ModMod-2 \\ \hline
		Google+ & $98.7\%$ & $\textbf{98.8\%}$ & $98.6\%$ \\
		LiveJournal & $72.9\%$ & $\textbf{75.8\%}$ & $74.3\%$ \\
	\end{tabular}
	\vspace{-0.2in}
\end{table}
We evaluate the quality of the seed sets returned by different algorithms using the techniques presented in Section~\ref{sec:analysis}. In our evaluation, we always choose the tighter bound between $\mu_3({X})$ and $\mu_4({X})$ as defined in \eqref{eq:ubounds}. We set $\delta=10^{-6}$ so that the approximation guarantees obtained by \eqref{eq:guarantee} have high confidence. Table~\ref{table:bounds} shows the approximation guarantees derived for the seed sets constructed with $2^{10}\times 10,000$ RR sets generated. As can be seen, the seed sets constructed by our algorithms have approximation guarantees above $98\%$ and $70\%$ for the Google+ and LiveJournal datasets respectively. This implies that (i) our proposed upper bounds on the maximum achievable profit are quite tight, and (ii) the heuristic algorithms perform rather close to the optimal.

\subsection{Normalization}\label{subsec:result_normalization}
\begin{figure}[!t]
	\capstart
	\centering
	\includegraphics[width=1.0\linewidth]{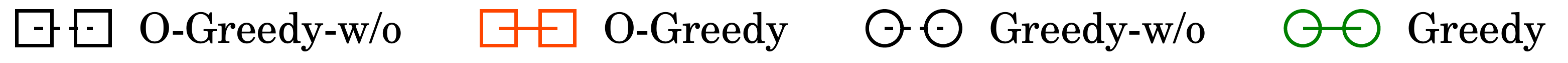}\vspace{-0.15in}\\
	\subfloat[Google+]{\includegraphics[width=0.48\linewidth]{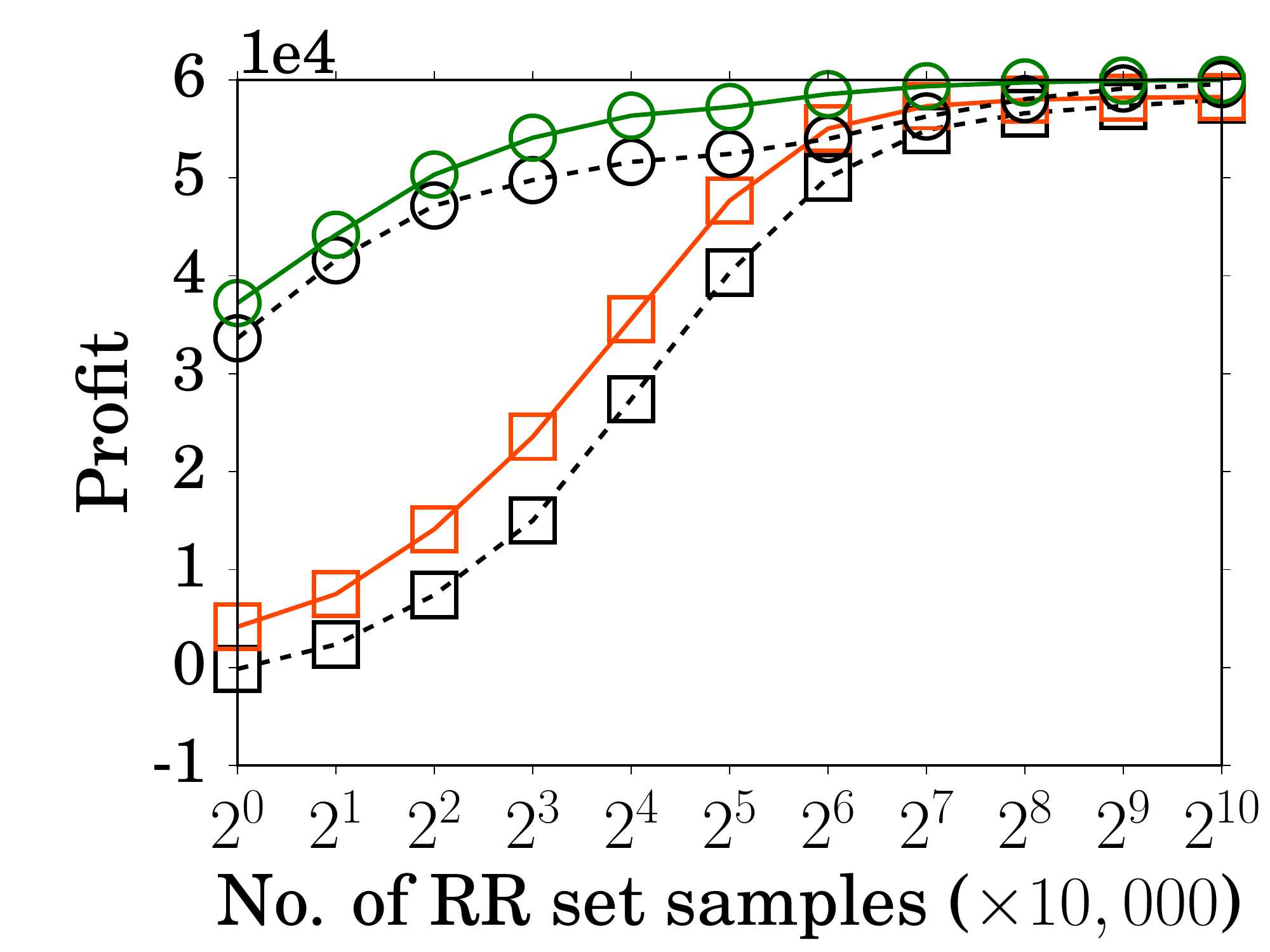}\label{subfig:gplus_profit_w_wo_norm}}\hfill
	\subfloat[LiveJournal]{\includegraphics[width=0.48\linewidth]{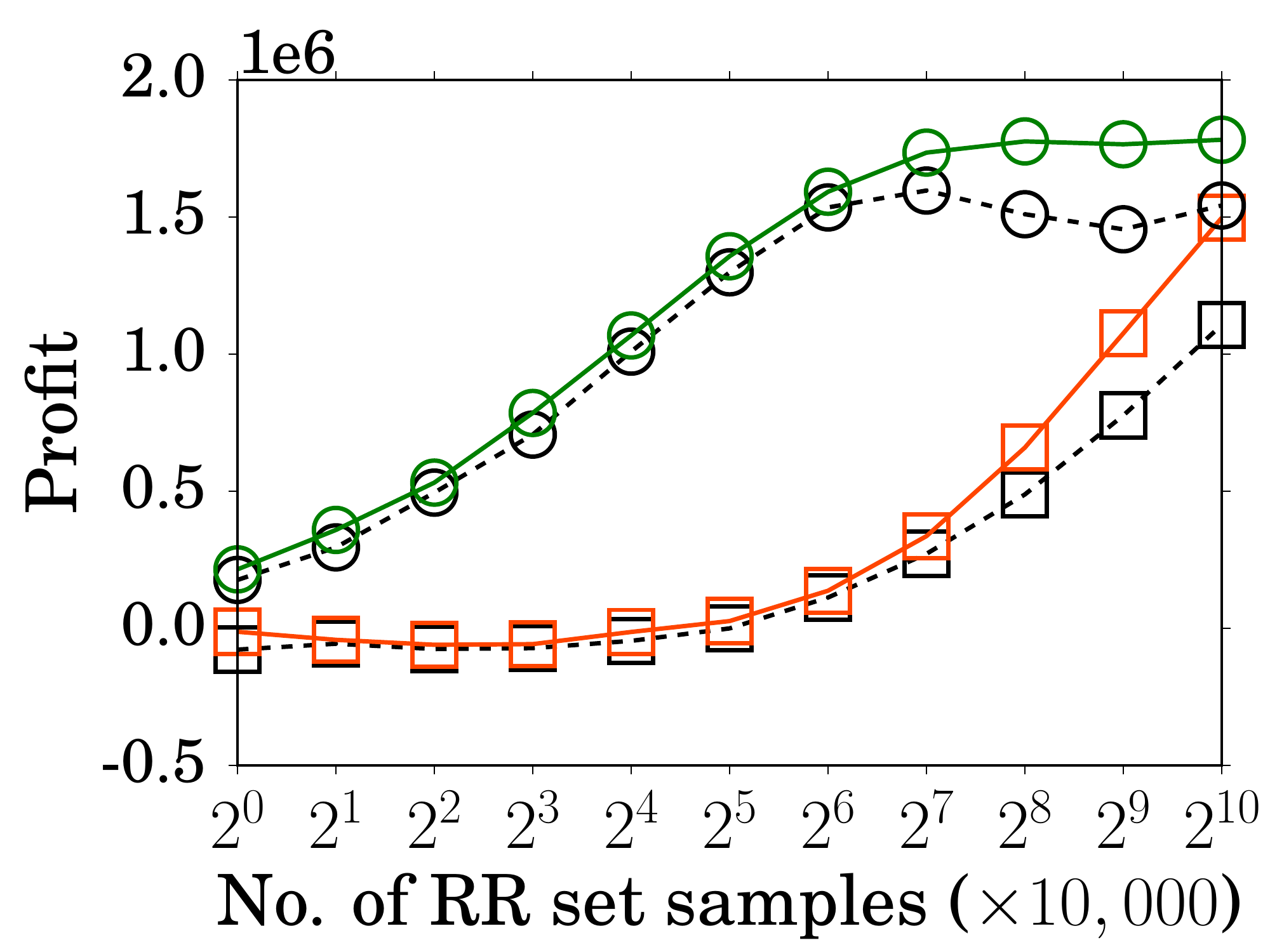}\label{subfig:liveJournal_profit_w_wo_norm}}\hfill
	\caption{Impact of normalization (algorithms with prefix ``O-'' do not use pruning technique and with postfix ``-w/o'' are without normalization).}\label{fig:profit_w_wo_norm}
	\vspace{-0.15in}
\end{figure}

In Section~\ref{subsec:result_prune}, we have shown that the normalization can increase the amount of search space reduction by the pruning technique. Now, we further evaluate the impact of normalization on the profit. \figurename~\ref{fig:profit_w_wo_norm} shows the profit produced by the Greedy algorithm with and without the normalization. The results for the ModMod algorithms are similar. As can be seen, no matter whether the pruning technique is used, the Greedy algorithm with normalization can always produce higher profit than that without normalization. This confirms Theorem~\ref{theorem:normalization} that the normalization can reduce the sampling error limit and thus, it can further improve the solution quality.

\section{Conclusion}\label{sec:conclusion}
In this paper, we have studied a profit maximization problem for OSN providers conducting viral marketing. The objective is to select initial seed nodes to maximize the total profit that accounts for both the benefit of influence spread and the cost of influence propagation. We have proposed a two-phase framework that first reduces the search space via an iterative pruning technique and then finds the solution via some heuristic algorithms. We have presented several bounds to measure the quality of the solution obtained by any algorithm. Experimental results with real OSN datasets demonstrate the effectiveness and efficiency of our techniques, and show the tightness of our derived upper bounds.

\section*{Acknowledgment}
This research is supported by Singapore Ministry of Education Academic Research Fund Tier 1 under Grant 2017-T1-002-024 and Tier 2 under Grant MOE2015-T2-2-114.

\bibliographystyle{IEEEtranS}
\bibliography{reference}

\begin{thebibliography}{10}
\providecommand{\url}[1]{#1}
\csname url@samestyle\endcsname
\providecommand{\newblock}{\relax}
\providecommand{\bibinfo}[2]{#2}
\providecommand{\BIBentrySTDinterwordspacing}{\spaceskip=0pt\relax}
\providecommand{\BIBentryALTinterwordstretchfactor}{4}
\providecommand{\BIBentryALTinterwordspacing}{\spaceskip=\fontdimen2\font plus
\BIBentryALTinterwordstretchfactor\fontdimen3\font minus
  \fontdimen4\font\relax}
\providecommand{\BIBforeignlanguage}[2]{{%
\expandafter\ifx\csname l@#1\endcsname\relax
\typeout{** WARNING: IEEEtranS.bst: No hyphenation pattern has been}%
\typeout{** loaded for the language `#1'. Using the pattern for}%
\typeout{** the default language instead.}%
\else
\language=\csname l@#1\endcsname
\fi
#2}}
\providecommand{\BIBdecl}{\relax}
\BIBdecl

\bibitem{Barbieri_TAI_2012}
N.~Barbieri, F.~Bonchi, and G.~Manco, ``Topic-aware social influence
  propagation models,'' in \emph{Proc. IEEE ICDM}, 2012, pp. 81--90.

\bibitem{Borgs_RIS_2014}
C.~Borgs, M.~Brautbar, J.~Chayes, and B.~Lucier, ``Maximizing social influence
  in nearly optimal time,'' in \emph{Proc. SODA}, 2014, pp. 946--957.

\bibitem{Chalermsook_SNM_2015}
P.~Chalermsook, A.~Das~Sarma, A.~Lall, and D.~Nanongkai, ``Social network
  monetization via sponsored viral marketing,'' in \emph{Proc. ACM SIGMETRICS},
  2015, pp. 259--270.

\bibitem{Chen_TAI_2015}
S.~Chen, J.~Fan, G.~Li, J.~Feng, K.-L. Tan, and J.~Tang, ``Online topic-aware
  influence maximization,'' \emph{Proc. VLDB Endowment}, vol.~8, no.~6, pp.
  666--677, 2015.

\bibitem{Chen_time_2012}
W.~Chen, W.~Lu, and N.~Zhang, ``Time-critical influence maximization in social
  networks with time-delayed diffusion process,'' in \emph{Proc. AAAI}, 2012,
  pp. 592--598.

\bibitem{Chen_MIA_2010}
W.~Chen, C.~Wang, and Y.~Wang, ``Scalable influence maximization for prevalent
  viral marketing in large-scale social networks,'' in \emph{Proc. ACM KDD},
  2010, pp. 1029--1038.

\bibitem{Chen_degreeDiscount_2009}
W.~Chen, Y.~Wang, and S.~Yang, ``Efficient influence maximization in social
  networks,'' in \emph{Proc. ACM KDD}, 2009, pp. 199--208.

\bibitem{Dagum_MCestimation_2000}
P.~Dagum, R.~Karp, M.~Luby, and S.~Ross, ``An optimal algorithm for monte carlo
  estimation,'' \emph{SIAM Journal on Computing}, vol.~29, no.~5, pp.
  1484--1496, 2000.

\bibitem{Domingos_maxInfluence_2001}
P.~Domingos and M.~Richardson, ``Mining the network value of customers,'' in
  \emph{Proc. ACM KDD}, 2001, pp. 57--66.

\bibitem{Du_CTIC_2013}
N.~Du, L.~Song, M.~Gomez-Rodriguez, and H.~Zha, ``Scalable influence estimation
  in continuous-time diffusion networks,'' in \emph{Proc. NIPS}, 2013, pp.
  3147--3155.

\bibitem{Fujishige_submodularOpt_2005}
S.~Fujishige, \emph{Submodular Functions and Optimization}.\hskip 1em plus
  0.5em minus 0.4em\relax Elsevier Science, 2005, vol.~58.

\bibitem{Goyal_infMax_2011}
A.~Goyal, F.~Bonchi, and L.~V.~S. Lakshmanan, ``A data-based approach to social
  influence maximization,'' \emph{Proc. VLDB Endowment}, vol.~5, no.~1, pp.
  73--84, 2011.

\bibitem{Iyer_submodularDiff_2012}
R.~Iyer and J.~Bilmes, ``Algorithms for approximate minimization of the
  difference between submodular functions, with applications,'' in \emph{Proc.
  UAI}, 2012, pp. 407--417.

\bibitem{Jung_IRIE_2012}
K.~Jung, W.~Heo, and W.~Chen, ``{IRIE}: Scalable and robust influence
  maximization in social networks,'' in \emph{Proc. IEEE ICDM}, 2012, pp.
  918--923.

\bibitem{Kempe_maxInfluence_2003}
D.~Kempe, J.~Kleinberg, and E.~Tardos, ``Maximizing the spread of influence
  through a social network,'' in \emph{Proc. ACM KDD}, 2003, pp. 137--146.

\bibitem{Khan_revmax_2016}
A.~Khan, B.~Zehnder, and D.~Kossmann, ``Revenue maximization by viral
  marketing: A social network host's perspective,'' in \emph{Proc. IEEE ICDE},
  2016.

\bibitem{Leskovec_CELF_2007}
J.~Leskovec, A.~Krause, C.~Guestrin, C.~Faloutsos, J.~VanBriesen, and
  N.~Glance, ``Cost-effective outbreak detection in networks,'' in \emph{Proc.
  ACM KDD}, 2007, pp. 420--429.

\bibitem{Leskovec_SNAP_2014}
J.~Leskovec and A.~Krevl, ``{SNAP Datasets}: {Stanford} large network dataset
  collection,'' \url{http://snap.stanford.edu/data}, 2014.

\bibitem{Lu_maxprofit_2012}
W.~Lu and L.~V. Lakshmanan, ``Profit maximization over social networks,'' in
  \emph{Proc. IEEE ICDM}, 2012, pp. 479--488.

\bibitem{Meyer_VVM_2015}
A.~Meyer, ``Viral video marketing: What's, why's \& how's of going viral,''
  \url{http://www.marketergizmo.com/viral-video-marketing-cats-babies-and-your-company/},
  2015.

\bibitem{Nemhauser_submodular_1978}
G.~L. Nemhauser, L.~A. Wolsey, and M.~L. Fisher, ``An analysis of
  approximations for maximizing submodular set functions-{I},''
  \emph{Mathematical Programming}, vol.~14, no.~1, pp. 265--294, 1978.

\bibitem{Nguyen_BCT_2016}
H.~T. Nguyen, T.~N. Dinh, and M.~T. Thai, ``Cost-aware targeted viral marketing
  in billion-scale networks,'' in \emph{Proc. IEEE INFOCOM}, 2016.

\bibitem{Nguyen_DSSA_2016}
H.~T. Nguyen, M.~T. Thai, and T.~N. Dinh, ``Stop-and-stare: Optimal sampling
  algorithms for viral marketing in billion-scale networks,'' in \emph{Proc.
  ACM SIGMOD}, 2016, pp. 695--710.

\bibitem{Ohsaka_prunedMC_2014}
N.~Ohsaka, T.~Akiba, Y.~Yoshida, and K.~Kawarabayashi, ``Fast and accurate
  influence maximization on large networks with pruned monte-carlo
  simulations,'' in \emph{Proc. AAAI}, 2014, pp. 138--144.

\bibitem{Roberts_socialAds_2016}
J.~J. Roberts, ``{Facebook and Google} are big winners as political ad money
  moves online,'' \emph{Fortune}, 2016.

\bibitem{Rodriguez_CT_2011}
M.~G. Rodriguez, D.~Balduzzi, and B.~Sch{\"o}lkopf, ``Uncovering the temporal
  dynamics of diffusion networks,'' in \emph{Proc. ICML}, 2011, pp. 561--568.

\bibitem{Song_CGA_2015}
G.~Song, X.~Zhou, Y.~Wang, and K.~Xie, ``Influence maximization on large-scale
  mobile social network: A divide-and-conquer method,'' \emph{IEEE Trans.
  Parallel and Distributed Systems}, vol.~26, no.~5, pp. 1379--1392, 2015.

\bibitem{Tang_OPIM_2018}
J.~Tang, X.~Tang, X.~Xiao, and J.~Yuan, ``Online processing algorithms for
  influence maximization,'' in \emph{Proc. ACM SIGMOD}, 2018.

\bibitem{Tang_profitMax_2016}
J.~Tang, X.~Tang, and J.~Yuan, ``Profit maximization for viral marketing in
  online social networks,'' in \emph{Proc. IEEE ICNP}, 2016, pp. 1--10.

\bibitem{Tang_infMax_2017}
J.~Tang, X.~Tang, and J.~{Yuan}, ``Influence maximization meets efficiency and
  effectiveness: A hop-based approach,'' in \emph{Proc. IEEE/ACM ASONAM}, 2017,
  pp. 64--71.

\bibitem{Tang_profitMaxUS_2018}
J.~Tang, X.~{Tang}, and J.~Yuan, ``Profit maximization for viral marketing in
  online social networks: Algorithms and analysis,'' \emph{IEEE Transactions on
  Knowledge and Data Engineering}, 2018.

\bibitem{Tang_IMM_2015}
Y.~Tang, Y.~Shi, and X.~Xiao, ``Influence maximization in near-linear time: A
  martingale approach,'' in \emph{Proc. ACM SIGMOD}, 2015, pp. 1539--1554.

\bibitem{Tang_reverse_2014}
Y.~Tang, X.~Xiao, and Y.~Shi, ``Influence maximization: Near-optimal time
  complexity meets practical efficiency,'' in \emph{Proc. ACM SIGMOD}, 2014,
  pp. 75--86.

\bibitem{Zhou_UBLF_2013}
C.~Zhou, P.~Zhang, J.~Guo, X.~Zhu, and L.~Guo, ``{UBLF}: An upper bound based
  approach to discover influential nodes in social networks,'' in \emph{Proc.
  IEEE ICDM}, 2013, pp. 907--916.

\bibitem{Zhu_maxprofit_2013}
Y.~Zhu, Z.~Lu, Y.~Bi, W.~Wu, Y.~Jiang, and D.~Li, ``Influence and profit: Two
  sides of the coin,'' in \emph{Proc. IEEE ICDM}, 2013, pp. 1301--1306.

\end{thebibliography}

\section*{APPENDIX}
\begin{lemma}\label{lemma:prune}
	${A}_{t}\subseteq{A}_{t+1}\subseteq{B}_{t+1}\subseteq{B}_{t}$ for any $t\geq 0$.
\end{lemma}
\begin{IEEEproof}[Proof of Lemma~\ref{lemma:prune}]
	By the definitions in lines~\ref{alg:IterativePrune_At} and \ref{alg:IterativePrune_Bt} of Algorithm~\ref{alg:IterativePrune}, ${A}_{t}$ gradually expands over iterations and ${B}_{t}$ gradually shrinks, i.e., ${A}_{t}\subseteq{A}_{t+1}$ and ${B}_{t+1}\subseteq{B}_{t}$. What is left is to show that ${A}_t$ is always a subset of ${B}_t$. We prove it by induction. Obviously, ${A}_{0}=\emptyset\subseteq{V}={B}_{0}$. Suppose that ${A}_{t}\subseteq{B}_{t}$ holds for some $t\geq 0$. Then, we can easily get that ${A}_t\subseteq{B}_{t+1}$ by the definition in line~\ref{alg:IterativePrune_Bt} of Algorithm~\ref{alg:IterativePrune} and that ${A}_{t+1}\subseteq{B}_t$ by the definition in line~\ref{alg:IterativePrune_At} of Algorithm~\ref{alg:IterativePrune}. On the other hand, for any node $v\in{B}_t\setminus{A}_t$, it is obvious that ${A}_t\subseteq{B}_t\setminus\{v\}$. Thus, for any node $v\in{A}_{t+1}\setminus{A}_t\subseteq{B}_t\setminus{A}_t$, we have $\beta(v\mid{A}_t)-\gamma(v\mid{B}_t\setminus\{v\})\geq \beta(v\mid{B}_t\setminus\{v\})-\gamma(v\mid{A}_t)>0$, where the first inequality is due to the submodularity of $\beta(\cdot)$ and $\gamma(\cdot)$ and the second inequality is by the definition of ${A}_{t+1}$ in line~\ref{alg:IterativePrune_At} of Algorithm~\ref{alg:IterativePrune}. It indicates that $v\in{B}_{t+1}$ as well. Therefore, ${A}_{t+1}\setminus{A}_t\subseteq{B}_{t+1}$, which implies ${A}_{t+1}\subseteq{B}_{t+1}$ since we already know that ${A}_t\subseteq{B}_{t+1}$. Thus, we can conclude that ${A}_{t+1}\subseteq{B}_{t+1}$.
\end{IEEEproof}

\begin{IEEEproof}[Proof of Theorem~\ref{theorem:pruneImprovement}]
	By the definition of ${S}_t$, we easily have ${A}_t\subseteq{S}_t\subseteq{B}_t$. We already know that ${A}_{t}\subseteq{A}_{t+1}\subseteq{B}_{t+1}\subseteq{B}_{t}$ by Lemma~\ref{lemma:prune}. Thus,
	\allowdisplaybreaks[4]
	\begin{align}
	{S}_{t+1}
	&={S}\cap{B}_{t+1}\cup{A}_{t+1}\notag\\
	&={S}\cap\big({B}_t\setminus({B}_t\setminus{B}_{t+1})\big)\cup{A}_{t+1}\notag\\
	&={S}\cap{B}_t\setminus\big({S}\cap({B}_t\setminus{B}_{t+1})\big)\cup{A}_{t+1}\notag\\
	&={S}\cap{B}_t\cup{A}_{t+1}\setminus\big({S}\cap({B}_t\setminus{B}_{t+1})\setminus{A}_{t+1}\big)\notag\\
	&={S}\cap{B}_t\cup{A}_{t+1}\setminus\big({S}\cap({B}_t\setminus{B}_{t+1})\big)\notag\\
	&={S}\cap{B}_t\cup{A}_t\cup({A}_{t+1}\setminus{A}_t)\setminus\big({S}\cap({B}_t\setminus{B}_{t+1})\big)\notag\\
	&={S}_{t}\cup({A}_{t+1}\setminus{A}_t)\setminus\big({S}\cap({B}_t\setminus{B}_{t+1})\big).\label{eq:S_t}
	\end{align}
	By line~\ref{alg:IterativePrune_At} of Algorithm~\ref{alg:IterativePrune}, for any node $v\in{A}_{t+1}\setminus{A}_t$, it holds that $\beta(v\mid{B}_t\setminus\{v\})-\gamma(v\mid{A}_t)>0$. Then, for any ${S}_t\subseteq{T}\subseteq{S}_t\cup({A}_{t+1}\setminus{A}_t)$, since ${A}_t\subseteq{T}$ and ${T}\subseteq{B}_t\setminus\{v\}$, we have $\phi(v\mid{T})=\beta(v\mid{T})-\gamma(v\mid{T})\geq\beta(v\mid{B}_t\setminus\{v\})-\gamma(v\mid{A}_t)>0$ due to the submodularity. Let $v_1,v_2,\dots,v_k$ be the set of nodes in ${A}_{t+1}\setminus{A}_t\setminus{S}_t$. It follows that
	\begin{align*}
	\phi({S}_t)
	&<\phi({S}_t\cup\{v_1\})\\
	&<\phi({S}_t\cup\{v_1,v_2\})\\
	&<\cdots\\
	&<\phi({S}_t\cup\{v_1,v_2,\dots,v_k\})\\
	&=\phi\big({S}_t\cup({A}_{t+1}\setminus{A}_t\setminus{S}_t)\big)\\
	&=\phi\big({S}_t\cup({A}_{t+1}\setminus{A}_t)\big).
	\end{align*}
	Similarly, by line~\ref{alg:IterativePrune_Bt} of Algorithm~\ref{alg:IterativePrune}, for any node $v\in{B}_t\setminus{B}_{t+1}$, it holds that $\beta(v\mid{A}_t)-\gamma(v\mid{B}_t\setminus\{v\})<0$. Then, for any ${S}_{t}\cup({A}_{t+1}\setminus{A}_t)\setminus\big({S}\cap({B}_t\setminus{B}_{t+1})\big)\subseteq{T}\subseteq{S}_t\cup({A}_{t+1}\setminus{A}_t)$, since ${A}_t\subseteq{T}$ and ${T}\subseteq{B}_t\setminus\{v\}$, we have $\phi(v\mid{T})=\beta(v\mid{T})-\gamma(v\mid{T})\leq\beta(v\mid{A}_t)-\gamma(v\mid{B}_t\setminus\{v\})<0$ due to the submodularity. Let $u_1,u_2,\cdots,u_l$ be the set of nodes in ${S}\cap({B}_t\setminus{B}_{t+1})$. It follows that
	\begin{align*}
	&\!\!\!\!\phi({S}_t\cup({A}_{t+1}\setminus{A}_t))\\
	&<\phi({S}_t\cup({A}_{t+1}\setminus{A}_t)\setminus\{u_1\})\\
	&<\phi({S}_t\cup({A}_{t+1}\setminus{A}_t)\setminus\{u_1,u_2\})\\
	&<\cdots\\
	&<\phi({S}_t\cup({A}_{t+1}\setminus{A}_t)\setminus\{u_1,u_2,\dots,u_l\})\\
	&=\phi\Big({S}_{t}\cup({A}_{t+1}\setminus{A}_t)\setminus\big({S}\cap({B}_t\setminus{B}_{t+1})\big)\Big)\\
	&=\phi({S}_{t+1}),
	\end{align*}
	where the last equality is due to \eqref{eq:S_t}. Therefore, $\phi({S}_t)<\phi({S}_{t+1})$ if ${A}_{t+1}\setminus{A}_t\setminus{S}_t\neq\emptyset$ or ${S}\cap({B}_t\setminus{B}_{t+1})\neq\emptyset$. On the other hand, if ${A}_{t+1}\setminus{A}_t\setminus{S}_t={S}\cap({B}_t\setminus{B}_{t+1})=\emptyset$, based on \eqref{eq:S_t}, we have ${S}_t={S}_{t+1}$.
\end{IEEEproof}

\begin{IEEEproof}[Proof of Corollary \ref{corollary:pruneImprovement}]
	According to Theorem~\ref{theorem:pruneImprovement}, $\phi({S})=\phi({S}_0)\leq\phi({S}_1)\leq\cdots\leq\phi({S}\cap{B}^\ast\cup{A}^\ast)$ since ${S}={S}\cap{B}_0\cup{A}_0={S}_0$. On the other hand, if every ``='' holds, we have ${S}={S}_0={S}_1=\cdots={S}\cap{B}^\ast\cup{A}^\ast$, which implies ${A}^\ast\subseteq{S}\subseteq{B}^\ast$. This is contradictory to ${S}\notin[{A}^\ast,{B}^\ast]$. Thus, it holds that $\phi({S})<\phi({S}\cap{B}^\ast\cup{A}^\ast)$.
\end{IEEEproof}

\begin{IEEEproof}[Proof of Corollary \ref{corollary:allMaximizers}]
	Suppose ${S}^\ast\notin[{A}^\ast,{B}^\ast]$. By Corollary~\ref{corollary:pruneImprovement}, we have $\phi({S}^\ast)<\phi({S}^\ast\cap{B}^\ast\cup{A}^\ast)$. This is contradictory to the fact that ${S}^\ast$ produces the maximum achievable profit. Thus, it holds that ${A}^\ast\subseteq{S}^\ast\subseteq{B}^\ast$.
\end{IEEEproof}

\begin{IEEEproof}[Proof of Theorem \ref{theorem:normalization_prune}]
	Let $\alpha({S})$ denote the overlapping value between the benefit of influence spread and the cost of influence propagation for a seed set ${S}$, which is defined as
	\begin{equation*}
	\alpha({S})=\mathbb{E}\Big[\sum_{v\in{V}_X({S})}\min\{b(v),c(v)\}\Big].
	\end{equation*}
	It is easy to verify that $\alpha(\cdot)$ is also submodular and monotone. By the definition of normalization in Section~\ref{subsec:problem}, we have $\beta({S})=\bar{\beta}({S})+\alpha({S})$ and $\gamma({S})=\bar{\gamma}({S})+\alpha({S})$. We prove that ${A}_t\subseteq\bar{{A}}_t\subseteq\bar{{B}}_t\subseteq{B}_t$ for any $t\geq 0$ by induction. Obviously, it holds that $\emptyset={A}_0=\bar{{A}}_0\subseteq\bar{{B}}_0={B}_0={V}$. Suppose that ${A}_t\subseteq\bar{{A}}_t\subseteq\bar{{B}}_t\subseteq{B}_t$ holds for some $t\geq 0$. 
	
	Then, for any node $v\in{B}_{t}\setminus\bar{{B}}_t$, we have $\beta(v\mid{B}_t\setminus\{v\})-\gamma(v\mid{A}_t)=\bar{\beta}(v\mid{B}_t\setminus\{v\})-\bar{\gamma}(v\mid{A}_t)+\alpha(v\mid{B}_t\setminus\{v\})-\alpha(v\mid{A}_t)\leq \bar{\beta}(v\mid{B}_t\setminus\{v\})-\bar{\gamma}(v\mid{A}_t)\leq \bar{\beta}(v\mid\bar{{B}}_t\setminus\{v\})-\bar{\gamma}(v\mid\bar{{A}}_t)$, where the inequalities are due to the submodularity of $\alpha(\cdot)$, $\bar{\beta}(\cdot)$ and $\bar{\gamma}(\cdot)$. By Lemma~\ref{lemma:prune}, $\bar{{A}}_{t-1}\subseteq\bar{{A}}_t\subseteq\bar{{B}}_{t-1}=\bar{{B}}_{t-1}\setminus\{v\}\subseteq\bar{{B}}_t\setminus\{v\}$. It follows from the submodularity that $\bar{\beta}(v\mid\bar{{B}}_t\setminus\{v\})-\bar{\gamma}(v\mid\bar{{A}}_t)\leq \bar{\beta}(v\mid\bar{{A}}_{t-1})-\bar{\gamma}(v\mid\bar{{B}}_{t-1}\setminus\{v\})<0$, where the last inequality is due to $v\notin\bar{{B}}_t$. This implies that $({A}_{t+1}\setminus{A}_t)\cap({B}_t\setminus\bar{{B}}_t)=\emptyset$. Since ${A}_{t+1}\setminus{A}_t\subseteq{B}_t$,  we have ${A}_{t+1}\setminus{A}_t\subseteq\bar{{B}}_t$. Since ${A}_t\subseteq\bar{{B}}_t$, it follows that ${A}_{t+1}\subseteq \bar{{B}}_t$. Thus, for any node $v\in{A}_{t+1}\setminus\bar{{A}}_{t}\subseteq \bar{{B}}_t\setminus\bar{{A}}_{t}$, we have $\bar{\beta}(v\mid\bar{{B}}_t\setminus\{v\})-\bar{\gamma}(v\mid\bar{{A}}_t)\geq\bar{\beta}(v\mid{B}_t\setminus\{v\})-\bar{\gamma}(v\mid{A}_t)=\beta(v\mid{B}_t\setminus\{v\})-\gamma(v\mid{A}_t)-\alpha(v\mid{B}_t\setminus\{v\})+\alpha(v\mid{A}_t)\geq \beta(v\mid{B}_t\setminus\{v\})-\gamma(v\mid{A}_t)>0$, where the first two inequalities are due to the submodularity and the last inequality is due to $v\in{A}_{t+1}$. This implies that ${A}_{t+1}\setminus\bar{{A}}_{t}\subseteq\bar{{A}}_{t+1}$. Since $\bar{A}_t\subseteq\bar{A}_{t+1}$, it follows that ${A}_{t+1}\subseteq\bar{{A}}_{t+1}$.
	
	Similarly, for any node $v\in\bar{{A}}_{t}\setminus{A}_t$, we have $\beta(v\mid{A}_t)-\gamma(v\mid{B}_t\setminus\{v\})=\bar{\beta}(v\mid{A}_t)-\bar{\gamma}(v\mid{B}_t\setminus\{v\})+\alpha(v\mid{A}_t)-\alpha(v\mid{B}_t\setminus\{v\})\geq \bar{\beta}(v\mid{A}_t)-\bar{\gamma}(v\mid{B}_t\setminus\{v\})\geq \bar{\beta}(v\mid\bar{{A}}_t)-\bar{\gamma}(v\mid\bar{{B}}_t\setminus\{v\})\geq \bar{\beta}(v\mid\bar{{B}}_{t-1}\setminus\{v\})-\bar{\gamma}(v\mid\bar{{A}}_{t-1})>0$, where the first three inequalities are due to the submodularity of $\alpha(\cdot)$, $\bar{\beta}(\cdot)$ and $\bar{\gamma}(\cdot)$, and the last inequality is due to $v\in\bar{{A}}_t$. This implies that $\bar{{A}}_t\setminus{A}_t\subseteq{B}_{t+1}$. Since ${A}_t\subseteq{A}_{t+1}\subseteq{B}_{t+1}$, it follows that $\bar{{A}}_{t}\subseteq {B}_{t+1}$. Thus, for any node $v\in\bar{{B}}_{t}\setminus{B}_{t+1}\subseteq \bar{{B}}_t\setminus\bar{{A}}_{t}$, we have $\bar{\beta}(v\mid\bar{{A}}_t)-\bar{\gamma}(v\mid\bar{{B}}_t\setminus\{v\})\leq\bar{\beta}(v\mid{A}_t)-\bar{\gamma}(v\mid{B}_t\setminus\{v\})=\beta(v\mid{A}_t)-\gamma(v\mid{B}_t\setminus\{v\})-\alpha(v\mid{A}_t)+\alpha(v\mid{B}_t\setminus\{v\})\leq \beta(v\mid{A}_t)-\gamma(v\mid{B}_t\setminus\{v\})<0$, where the first two inequalities are due to the submodularity and the last inequality is due to $v\notin{B}_{t+1}$. This implies that $\bar{{B}}_{t+1}\cap(\bar{{B}}_{t}\setminus{B}_{t+1})=\emptyset$. Since $\bar{{B}}_{t+1}\subseteq\bar{{B}}_t$, it follows that $\bar{{B}}_{t+1}\subseteq{B}_{t+1}$. 
	By induction, we have ${A}_t\subseteq\bar{{A}}_t\subseteq\bar{{B}}_t\subseteq{B}_t$ for any $t\geq 0$ (it also holds after converged) and thus, ${A}^\ast\subseteq\bar{{A}}^\ast\subseteq\bar{{B}}^\ast\subseteq{B}^\ast$.
\end{IEEEproof}

\begin{IEEEproof}[Proof of Theorem \ref{theorem:modularGlobalBound}]
	Let ${S}^\ast$ be an optimal seed set producing the maximum achievable profit. We can directly obtain that $\mu_i({X})\geq m_{X}^i({S}^\ast;\beta)-h_{X}^{\pi^\ast}({S}^\ast;\gamma)\geq\beta({S}^\ast)-\gamma({S}^\ast)=\phi({S}^\ast)$, where the first inequality is by the definition of $\mu_i({X})$ and the second inequality is due to the modular upper and lower bounds.
\end{IEEEproof}

\begin{IEEEproof}[Proof of Theorem \ref{theorem:sample_error}]
	Let $\lambda_\beta({S})$ be the expected fraction of samples covered by the seed set ${S}$. We have $\lambda_\beta({S})=\beta({S})/B$. To simplify the notation, we omit the common symbol ${S}$ in what follows, e.g., $\beta$ represents $\beta({S})$ and $\Lambda_\beta$ represents $\Lambda_\beta({S})$. Then, the inequalities to prove are equivalent to \[\Pr\left[\lambda_\beta<\left(\sqrt{\Lambda_\beta+0.25a} - 0.5\sqrt{a}\right)^2/\theta_\beta\right]\leq \delta/2,\] and \[\Pr\left[\lambda_\beta>\left(\sqrt{\Lambda_\beta+0.25a} + 0.5\sqrt{a}\right)^2/\theta_\beta\right]\leq \delta/2.\] 
	
	We prove the former first. In fact,
	\begin{align*}
	&\Pr\left[\lambda_\beta<\left(\sqrt{\Lambda_\beta+0.25a} - 0.5\sqrt{a}\right)^2/\theta_\beta\right]\\
	&\quad=\Pr\left[\sqrt{\lambda_\beta\theta_\beta}<\sqrt{\Lambda_\beta+0.25a} - 0.5\sqrt{a}\right]\\
	&\quad=\Pr\left[\left(\sqrt{\lambda_\beta\theta_\beta}+0.5\sqrt{a}\right)^2<\Lambda_\beta+0.25a\right]\\
	&\quad=\Pr\left[\Lambda_\beta-\lambda_\beta\theta_\beta>\sqrt{a\lambda_\beta\theta_\beta}\right]\\
	&\quad\leq \exp\left(-\frac{a\lambda_\beta\theta_\beta}{4(e-2)\lambda_\beta\theta_\beta}\right)\\
	&\quad=\delta/2,
	\end{align*}
	where the inequality is due to Lemma~\ref{lemma:Chernoff}.
	
	The proof of the latter is analogous.
	\begin{align*}
	&\Pr\left[\lambda_\beta>\left(\sqrt{\Lambda_\beta+0.25a} + 0.5\sqrt{a}\right)^2/\theta_\beta\right]\\
	&\quad=\Pr\left[\sqrt{\lambda_\beta\theta_\beta}>\sqrt{\Lambda_\beta+0.25a} + 0.5\sqrt{a}\right]\\
	&\quad=\Pr\left[\left(\sqrt{\lambda_\beta\theta_\beta}-0.5\sqrt{a}\right)^2>\Lambda_\beta+0.25a\right]\\
	&\quad=\Pr\left[\Lambda_\beta-\lambda_\beta\theta_\beta<-\sqrt{a\lambda_\beta\theta_\beta}\right]\\
	&\quad\leq \exp\left(-\frac{a\lambda_\beta\theta_\beta}{4(e-2)\lambda_\beta\theta_\beta}\right)\\
	&\quad=\delta/2
	\end{align*}
	
	Hence, the theorem is proven.
\end{IEEEproof}

\begin{IEEEproof}[Proof of Theorem \ref{theorem:ds_optimal_upper_bound}]
	Imagine that we evaluate ${S}^\ast$ with the benefit and cost samples. By Theorem~\ref{theorem:sample_error}, we know that
	\begin{align}
	&\!\!\!\!\Pr\left[\phi({S}^\ast)\leq \beta_{u}({S}^\ast)-\gamma_{l}({S}^\ast)\right]\nonumber\\
	&\geq \Pr\left[\big(\beta({S}^\ast)\leq\beta_{u}({S}^\ast)\big)\wedge\big(\gamma({S}^\ast)\geq\gamma_{l}({S}^\ast)\big)\right]\nonumber\\
	&\geq 1-\Big(\Pr\left[\beta({S}^\ast)> \beta_{u}({S}^\ast)\right]+ \Pr\left[\gamma({S}^\ast)< \gamma_{l}({S}^\ast)\right]\Big)\nonumber\\
	&\geq1-\left(\frac{\delta}{2}+\frac{\delta}{2}\right)\nonumber\\
	&=1-\delta\label{eq:ds_upper_bound_optimal1}.
	\end{align} 
	
	However, without knowing ${S}^\ast$, $\beta_u({S}^\ast)$ and $\gamma_l({S}^\ast)$ cannot be obtained. In what follows, we are going to bound $\beta_{u}({S}^\ast)-\gamma_{l}({S}^\ast)$ by an upper bound on the function $\eta({S})=\beta_{u}({S})-\gamma_{l}({S})$. To simplify the notations, and we omit the common symbol ${S}$ in what follows, e.g., $\eta$ represents $\eta({S})$. Then, the estimated profit $\tilde{\phi}=\rho_\beta\Lambda_\beta-\rho_\gamma\Lambda_\gamma$. Consider $\eta$ as a function of $\tilde{\phi}$ and $\Lambda_\beta$. We study the monotonicity of $\eta$ with respect to $\tilde{\phi}$ and $\Lambda_\beta$. By the definition of $\eta$, we have
	\begin{align*}
	\eta(\Lambda_\beta,\tilde{\phi})
	&=\rho_\beta\left(\Lambda_\beta+\sqrt{a(\Lambda_\beta+0.25a)}+0.5a\right)\\
	&\ \ \ -\rho_\gamma\left(\Lambda_\gamma-\sqrt{a(\Lambda_\gamma+0.25a)}+0.5a\right)\\
	&=\tilde{\phi}+\rho_\gamma\sqrt{a\big((\rho_\beta\Lambda_\beta-\tilde{\phi})/\rho_\gamma+0.25a\big)}\\
	&\ \ \ +0.5a(\rho_\beta-\rho_\gamma)+\rho_\beta\sqrt{a(\Lambda_\beta+0.25a)}.
	\end{align*}
	which is increasing with $\Lambda_\beta$ under any given $\tilde{\phi}$. Meanwhile, we also have
	\begin{align*}
	\eta(\Lambda_\beta,\tilde{\phi})
	&=\rho_\beta\left(\sqrt{\Lambda_\beta+0.25a}+0.5\sqrt{a}\right)^2\\
	&\ \ \ -\rho_\gamma\left(\sqrt{\Lambda_\gamma+0.25a}-0.5\sqrt{a}\right)^2\\
	&=\rho_\beta\left(\sqrt{\Lambda_\beta+0.25a}+0.5\sqrt{a}\right)^2\\
	&\ \ \ -\rho_\gamma\left(\sqrt{(\rho_\beta\Lambda_\beta-\tilde{\phi})/\rho_\gamma+0.25a}-0.5\sqrt{a}\right)^2,
	\end{align*}
	which is increasing with $\tilde{\phi}$ under any given $\Lambda_\beta$.
	
	For an estimated upper bound $\tilde{\mu}({S}^o)$ on the maximum achievable profit, we have $\tilde{\phi}({S}^\ast)\leq \tilde{\mu}({S}^o)$. On the other hand, it naturally holds that $\Lambda_\beta({S}^\ast)\leq \theta_\beta$. Thus, 
	\begin{align*}
	&\!\!\!\beta_{u}({S}^\ast)-\gamma_{l}({S}^\ast)\\
	&=\eta\big(\Lambda_\beta({S}^\ast), \tilde{\phi}({S}^\ast)\big)\\
	&\leq \eta\big(\theta_\beta, \tilde{\mu}({S}^o)\big)\\
	&=\tilde{\mu}({S}^o)+\rho_\gamma\sqrt{a\Big(\big(\rho_\beta\theta_\beta-\tilde{\mu}({S}^o)\big)/\rho_\gamma+0.25a\Big)}\\
	&\ \ \ +0.5a(\rho_\beta-\rho_\gamma)+\rho_\beta\sqrt{a(\theta_\beta+0.25a)}.
	\end{align*}
	Hence, the theorem is proven.
\end{IEEEproof}

\begin{IEEEproof}[Proof of Theorem \ref{theorem:normalization}]
	Consider the sampling error limit $\varepsilon$ under $\theta$ samples that can provide a probability guarantee of $1-\delta$, i.e., $\Pr[-\varepsilon\leq\Lambda-\lambda\theta\leq\varepsilon]\geq 1-\delta$ (where $\lambda$ is the expected value of the random variable and $\Lambda$ is the sum of $\theta$ samples). According to Lemma~\ref{lemma:Chernoff}, $\varepsilon$ is given by 
	\begin{equation}\label{eq:varepsilon}
	\varepsilon=\sqrt{4(e-2)\ln(2/\delta)\lambda\theta}=\sqrt{a\lambda\theta},
	\end{equation}
	where $a=4(e-2)\ln(2/\delta)$.
	
	We denote $\varepsilon_\beta$ and $\varepsilon_\gamma$ as the sampling error limits for the benefit and cost metrics under $\theta_\beta$ and $\theta_\gamma$ samples that can satisfy $\Pr[-\varepsilon_\beta\leq\tilde{\beta}({S})-\beta({S})\leq\varepsilon_\beta]\geq 1-\delta$ and $\Pr[-\varepsilon_\gamma\leq\tilde{\gamma}({S})-\gamma({S})\leq\varepsilon_\gamma]\geq 1-\delta$. Then, $\varepsilon_\phi=\varepsilon_\beta+\varepsilon_\gamma$ gives a total sampling error limit for the profit metric which guarantees $\Pr[-\varepsilon_\phi\leq\tilde{\phi}({S})-\phi({S})\leq\varepsilon_\phi]\geq 1-2\delta$. Likewise, let $\bar{\varepsilon}_\beta$, $\bar{\varepsilon}_\gamma$ and $\bar{\varepsilon}_\phi$ denote the sampling error limits under the normalized form. 
	
	Similar to the definitions of $\Upsilon_b$ and $\Upsilon_c$, we define $\bar{\Upsilon}_b=\bar{b}(V)=\sum_{v\in{V}}\bar{b}(v)$ and $\bar{\Upsilon}_c=\bar{c}(V)=\sum_{v\in{V}}\bar{c}(v)$. Then, $\bar{\Upsilon}_b-\bar{\Upsilon}_c=\Upsilon_b-\Upsilon_c$. For a given seed set ${S}$, let $\bar{\lambda}_\beta({S})$ and $\bar{\lambda}_\gamma({S})$ denote the expected fractions of benefit and cost samples covered by ${S}$ under the normalized form. To simplify the notation, we omit the common symbol ${S}$ in what follows, e.g., $\beta$ represents $\beta({S})$ and $\bar{\lambda}_\beta$ represents $\bar{\lambda}_\beta({S})$. Since $\beta=\frac{\Upsilon_b}{\theta_\beta}\cdot \lambda_\beta\theta_\beta$, based on the definition of $\varepsilon$ in \eqref{eq:varepsilon}, we have
	\begin{equation}\label{eq:varepsilon_beta}
	\varepsilon_\beta=\frac{\Upsilon_b}{\theta_\beta}\cdot\varepsilon=\frac{\Upsilon_b\sqrt{a\lambda_\beta\theta_\beta}}{\theta_\beta}=\sqrt{\frac{a\Upsilon_b^2\lambda_\beta}{\theta_\beta}}=\sqrt{\frac{a\Upsilon_b\beta}{\theta_\beta}}.
	\end{equation}
	Similarly, we have
	\begin{equation}
	\bar{\varepsilon}_\beta=\sqrt{\frac{a\bar{\Upsilon}_b\bar{\beta}}{\theta_\beta}},\quad\varepsilon_\gamma=\sqrt{\frac{a\Upsilon_c\gamma}{\theta_\gamma}} \quad\text{and}\quad\bar{\varepsilon}_\gamma=\sqrt{\frac{a\bar{\Upsilon}_c\bar{\gamma}}{\theta_\gamma}}.\label{eq:varepsilon_more}
	\end{equation}
	Note that $\Upsilon_b\geq \bar{\Upsilon}_b$, $\Upsilon_c\geq \bar{\Upsilon}_c$, $\beta\geq \bar{\beta}$ and $\gamma\geq \bar{\gamma}$. Together with \eqref{eq:varepsilon_beta} and \eqref{eq:varepsilon_more}, we have $\varepsilon_\beta\geq\bar{\varepsilon}_\beta$ and $\varepsilon_\gamma\geq\bar{\varepsilon}_\gamma$. Therefore, $\varepsilon_\phi=\varepsilon_\beta+\varepsilon_\gamma\geq\bar{\varepsilon}_\beta+\bar{\varepsilon}_\gamma=\bar{\varepsilon}_\phi$.
\end{IEEEproof}
\balance

\end{document}